\documentclass[times,double space]{qjrms4}

\usepackage[dvips,colorlinks,bookmarksopen,bookmarksnumbered,breaklinks=true,citecolor=red,urlcolor=red]{hyperref}

\usepackage{natbib}
\newcommand\BibTeX{{\rmfamily B\kern-.05em \textsc{i\kern-.025em b}\kern-.08em
T\kern-.1667em\lower.7ex\hbox{E}\kern-.125emX}}

\usepackage{moreverb}

\usepackage{amsmath} 
\usepackage{amssymb} 
\usepackage{amsfonts}
\usepackage{graphicx,color}                 
\usepackage{subfigure}
\usepackage[switch,pagewise,columnwise]{lineno}
\usepackage{color}
\usepackage[absolute,overlay]{textpos}
\usepackage{multirow}

\def\volumeyear{2012}

\begin{document}

\runningheads{X. Luo and I. Hoteit}{Efficient particle filtering through residual nudging}

\title{Efficient particle filtering through residual nudging}

\author{Xiaodong Luo$^1$\corrauth, and Ibrahim Hoteit$^2$}
\address{
$^1$International Research Institute of Stavanger, 5008 Bergen, Norway \\
$^2$MCSE, King Abdullah University of Science and Technology, Thuwal 23955-6900, Saudi Arabia
}

\corraddr{International Research Institute of Stavanger, 5008 Bergen, Norway. E-mail: xiaodong.luo@iris.no}

\date{}



\begin{abstract}
We introduce an auxiliary technique, called residual nudging, to the particle filter to enhance its performance in cases that it performs poorly. The main idea of residual nudging is to monitor, and if necessary, adjust the residual norm of a state estimate in the observation space so that it does not exceed a pre-specified threshold. We suggest a rule to choose the pre-specified threshold, and construct a state estimate accordingly to achieve this objective. Numerical experiments suggest that introducing residual nudging to a particle filter may (substantially) improve its performance, in terms of filter accuracy and/or stability against divergence, especially when the particle filter is implemented with a relatively small number of particles.
\end{abstract}


\maketitle

\section{Introduction} \label{sec:introduction}

State estimation often arises in geosciences studies. Recursive Bayesian filtering approaches, including the ensemble Kalman filter (EnKF, see, for examples, \citealt{Anderson-ensemble,Bishop-adaptive,Burgers-analysis,Hoteit2002,Whitaker-ensemble}) and the particle filter (PF, see, for examples, \citealt{Pham2001,vanLeeuwen-variance,vanLeeuwen2010-nonlinear}), are among the most popular data assimilation methods that are employed to tackle the problem. The EnKF and the PF provide approximations to the optimal solution obtained in the framework of recursive Bayesian estimation (RBE, see, for example, \citealp{Arulampalam2002}, or Section \ref{sec:ps}). The EnKF approximates the prior and posterior probability density functions (pdfs) of the model state by some Gaussian ones, which appears insufficient when the distribution of the model state is multi-modal\footnote{In such circumstances, it is more appropriate to use a mixture of Gaussian pdfs to approximate the distribution of the model state, see, for examples, \citet{Anderson-Monte,Bengtsson2003,Hoteit2008,Hoteit2012,Luo2008-spgsf1}.}. In contrast, the PF approximates the prior and posterior pdfs of the model state by mixture models of Dirac delta functions (i.e., Monte Carlo approximation), in which the mass points are particles drawn from certain pdfs. As the number of particles increases, the mixture models of Dirac delta functions approach the targeted pdfs asymptotically, hence the solution of the PF converges to the optimal one in the framework of RBE \cite[ch. 2]{Doucet2001-sequential}. The asymptotic convergence of the PF is achieved regardless of the presence of nonlinearity and non-Gaussianity in data assimilation.

A well-known problem in applying the PF is the phenomenon of weight collapse, also known as weight degeneracy or impoverishment (cf., for examples, \citealt{Arulampalam2002,Bengtsson2008,Gordon1993,Snyder2008}), in which the weight of a particular particle approaches one, and those associated with the remaining particles collapse to zero. In such circumstance, the effective sample size of the particle filter becomes very small, which often deteriorates the performance of the filter.

In the literature two strategies are often employed to tackle the problem of weight collapse \citep{Arulampalam2002}. One is to introduce a re-sampling step to the particle filter when the effective sample size is below a certain threshold. With the re-sampling step, a new set of particles is generated with identical weights. Many implementations of the particle filter differ from each other mainly at the re-sampling step, which, however, is a topic beyond the scope of this work. Readers are referred to, for examples, \citet{Arulampalam2002,VanLeeuwen2009}, for more information. A potential problem with the re-sampling strategy alone is that in certain circumstances, in order to avoid weight collapse, the number of particles may have to scale exponentially with the dimension of the model state \citep{Bengtsson2008,Snyder2008}. This implies that the PF may become prohibitively expensive for data assimilation in high-dimensional systems.

Another strategy is to choose a good importance (or proposal) density from which the particles are drawn \citep{Arulampalam2002,bocquet2010beyond,vanLeeuwen2010-nonlinear}. For instance, one may adopt an ``optimal'' importance density in the sense that, for a given particle at the current assimilation cycle, the weights of the samples drawn from the optimal importance density at the next assimilation cycle will be identical, regardless of the locations of the drawn samples \citep[Eq. (53)]{Arulampalam2002}. In a recent work, \citet{bocquet2010beyond} show that, in the 40-dimensional Lorenz 95 model (\citealt{Lorenz-optimal}, L95 hereafter), the particle filter equipped with the optimal importance density (in many cases substantially) outperforms the conventional bootstrap particle filter \citep{Gordon1993} when the sample size is no larger than 10000. A similar idea is also explored in \citet{vanLeeuwen2010-nonlinear,ades2012exploration}. There the authors adopt an importance density through which the generated particles are equipped with almost equal weights. By a proper design of the importance density, the PF with only 20 particles can achieve an estimation accuracy that is comparable to that of the conventional methods with thousands of particles \citep{vanLeeuwen2010-nonlinear}.

Apart from weight degeneracy, another factor that also influences the practical performance of the PF is the slow convergence rate of Monte Carlo approximation. After all, in many real-world problems, one can only afford to run finitely many -- often a small number of -- particles with the timing and computational cost limitations. In such circumstances, the slow convergence rate of Monte Carlo approximation implies that a PF solution with only finitely many particles may not be able to converge sufficiently close to the optimal one, and that, in this specific context, it may become an unrealistic objective for one to achieve the asymptotic optimality of the PF. A certain gap might arise between the optimal solution and the approximate one of the PF, especially when the sample size of the PF is relatively small. In this regard, introducing a re-sampling step to the PF alone may not be sufficient to address the effect of finite sample size. Instead, one may opt to seek some auxiliary technique to enhance the performance of the PF with a finite sample size, which is the focus of this study.

In this work we consider one possible auxiliary technique, called residual nudging, which aims to provide certain compensation to the PF solution when the filter does not perform well. Here a ``residual'' is a vector in the observation space, and is defined as the projection of a state estimate onto the observation space subtracted by the corresponding observation. In residual nudging our objective is to make the (weighted) vector norm of the residual (``residual norm'' for short) no larger than a pre-specified value. This is motivated by the observation that, if the residual norm is too large, then the corresponding state estimate is often a poor one. In such cases, it is better off to choose as the new estimate a state vector whose residual norm is smaller. In this sense, residual nudging can be considered as a safeguard strategy that helps a poorly-performing filter to perform less poorly by providing certain compensation to the original mean estimate of the PF. It, however, may not in its own right reduce the sample size requirement in order for the PF to obtain a reasonable approximate solution in data assimilation. {\color{black}{Likewise, it neither solves the weight collapse (or degeneracy) problem in the PF}.}  

This study is organized as follows. Section \ref{sec:ps} introduces the problem of our interest and presents the recursive Bayesian estimation as the conceptual solution. Section \ref{sec:RPF_RN} reviews the main steps in the particle filter, introduces the concept of residual nudging, and discusses how residual nudging can be implemented in a particle filter. Section \ref{sec:example_scalar} examines and compares the performance of the regularized particle filter (as a representative of the various particle filters), and that of the regularized particle filter with residual nudging, in a linear scalar model. This example is used to examine the effect of residual nudging on the performance of the regularized particle filter, in case that the filter performance is reasonably good with a relatively large sample size. Section \ref{sec:example} examines and compares the performance of the above two filters with the L95 model in different scenarios. In some experiments the performance of the regularized particle filter may be less satisfactory, due to the effect of finite sample sizes. In such cases, we show that residual nudging can help to improve the filter performance in terms of filter accuracy and/or stability against divergence. Section \ref{sec:conclusion} summarizes the whole work and discusses possible extensions of the present study.

\section{Particle filtering} \label{sec:ps}

Consider the state estimation problem in the following system
\begin{linenomath*}
\begin{subequations} \label{ps}
\begin{align}
 \label{ps_dyanmical_system} & \mathbf{x}_k  = \mathcal{M}_{k,k-1} \left( \mathbf{x}_{k-1}\right) + \varepsilon_k^m \, ,  \\
  \label{ps_observation_system} &  \mathbf{y}_k  = \mathcal{H}_{k} \left( \mathbf{x}_{k}\right) + \varepsilon_k^o \, .
\end{align}
\end{subequations}
\end{linenomath*}
Here, $\mathbf{x}_{k} \in \mathbb{R}^{n}$ is the $n$-dimensional model state at time instant $k$, $\mathbf{y}_{k} \in \mathbb{R}^{p}$ the corresponding observation of $\mathbf{x}_{k}$, $\varepsilon_k^m \in \mathbb{R}^{n}$ the model error with zero mean and covariance matrix $\mathbf{Q}_k$, and $\varepsilon_k^o \in \mathbb{R}^{p}$ the observation noise with zero mean and covariance matrix $\mathbf{R}_k$. The transition operator $\mathcal{M}_{k,k-1}: \mathbb{R}^{n} \rightarrow \mathbb{R}^{n}$ maps $\mathbf{x}_{k-1}$ to $\mathbf{x}_{k}$, and the observation operator $\mathcal{H}_{k}: \mathbb{R}^{n} \rightarrow \mathbb{R}^{p}$ projects $\mathbf{x}_{k}$ from the state space onto the observation space. The problem of our interest is to estimate the posterior pdf of the model state $\mathbf{x}_k$ at time instant $k$, given the observations $\mathbf{Y}_k = \left \{ \mathbf{y}_k,  \mathbf{y}_{k-1}, \dotsb \right \} $ up to and including $k$, together with the prior pdf $p \left( \mathbf{x}_{i} | \mathbf{Y}_{i-1} \right)$ of the model state $\mathbf{x}_i$ at some earlier instant $i$ ($i \le k$). For convenience of discussion, we assume that $p \leq n$ throughout this work.

Recursive Bayesian estimation (RBE) \citep{Arulampalam2002} provides a probabilistic framework that recursively solves the state estimation problem in terms of some conditional pdfs. Let $p \left( \mathbf{x}_{k} | \mathbf{Y}_{k-1} \right)$ be the prior pdf of $\mathbf{x}_{k}$ conditioned on the observations $\mathbf{Y}_{k-1}$ up to and including time $k-1$, but without the knowledge of the observation $\mathbf{y}_{k}$ yet. Once the observation $\mathbf{y}_k$ is known, one incorporates the information content of $\mathbf{y}_k$ according to Bayes' rule to update the prior pdf to the posterior one $p \left( \mathbf{x}_{k} | \mathbf{Y}_{k} \right)$. By evolving $p \left( \mathbf{x}_{k} | \mathbf{Y}_{k} \right)$ forward in time, one obtains a prior pdf $p \left( \mathbf{x}_{k+1} | \mathbf{Y}_{k} \right)$ at the next time instant. Concretely, the mathematical description of RBE consists of \citep{Arulampalam2002}:\\
\noindent Prediction step:
\begin{linenomath*}
\begin{equation} \label{BRR:prediction}
p \left( \mathbf{x}_{k} | \mathbf{Y}_{k-1} \right) =  \int p \left( \mathbf{x}_{k} | \mathbf{x}_{k-1} \right)  p \left( \mathbf{x}_{k-1} | \mathbf{Y}_{k-1} \right) d\mathbf{x}_{k-1} \, ,
\end{equation} \\
\noindent and Filtering step:
\begin{linenomath*}
\begin{equation} \label{BRR:update}
p \left( \mathbf{x}_{k} | \mathbf{Y}_{k} \right) =  \dfrac{ p \left( \mathbf{y}_{k} | \mathbf{x}_{k}  \right) p \left( \mathbf{x}_{k} | \mathbf{Y}_{k-1} \right) }{\int p \left( \mathbf{y}_{k} | \mathbf{x}_{k}  \right) p \left( \mathbf{x}_{k} | \mathbf{Y}_{k-1} \right)  d\mathbf{x}_{k}} \, ,
\end{equation}
\end{linenomath*}
\end{linenomath*}
where the transition pdf $p \left( \mathbf{x}_{k} | \mathbf{x}_{k-1}  \right)$ and the likelihood function $p \left( \mathbf{y}_{k} | \mathbf{x}_{k}  \right)$ are assumed known, in light of the knowledge of the distributions of the model and observation errors in Eq.~(\ref{ps}). Once the explicit forms of the conditional pdfs in  Eqs.~(\ref{BRR:prediction}) and (\ref{BRR:update}) are obtained, the optimal estimate and other associated statistical information can be derived based on a certain optimality criterion, e.g., minimum variance or maximum likelihood. Thus RBE provides a solution of the estimation problem, and conceptually leads to the optimal nonlinear filter.

In practice, however, difficulties often arise in deriving the exact optimal filter, largely due to the fact that the integrals in Eqs.~(\ref{BRR:prediction}) and (\ref{BRR:update}) are often intractable. Therefore one may have to adopt a certain approximation scheme for evaluation. In the PF, Monte Carlo approximation is adopted to approximate the prior and posterior pdfs. For instance, the posterior $p \left( \mathbf{x}_{k-1} | \mathbf{Y}_{k-1} \right)$ at the $(k-1)$th step is approximated by
\begin{linenomath*}
\[
p \left( \mathbf{x}_{k-1} | \mathbf{Y}_{k-1} \right) \approx \sum\limits_{i=1}^N w_{k-1,i} \delta(\mathbf{x}_{k-1} - \mathbf{x}^a_{k-1,i}) \, ,
\]
\end{linenomath*}
where $\mathbf{x}^a_{k-1,i}$ ($i = 1, 2, \dotsb, N$) are the particles at the filtering step before a re-sampling algorithm (if necessary) is applied, $w_{k-1,i}$ are the associated weights, and $N$ is the total number of particles (called sample size hereafter). For notational convenience, let $\tilde{\mathbf{x}}^a_{k-1,i}$ be the particles generated by a re-sampling algorithm, and $\tilde{w}_{k-1,i}$ the corresponding weights \footnote{If there is in fact no need to conduct re-sampling, then $\tilde{\mathbf{x}}^a_{k-1,i} = \mathbf{x}^a_{k-1,i}$ and $\tilde{w}_{k-1,i} = w_{k-1,i}$. If re-sampling is conducted, then $\tilde{w}_{k-1,i} = 1/N$, and $\tilde{\mathbf{x}}^a_{k-1,i}$ may be different from $\mathbf{x}^a_{k-1,i}$.}. In consistency with  Eqs.~(\ref{BRR:prediction}) and (\ref{BRR:update}), the PF has the following steps.

\underline{Prediction step:} The particles $\tilde{\mathbf{x}}^a_{k-1,i}$ are integrated forward with the
model to obtain the propagations $\mathbf{x}^b_{k,i}$ at the next time instant $k$. The associated weights of the new particles $\mathbf{x}^b_{k,i}$ remain to be $\tilde{w}_{k-1,i}$. This is equivalent to using the transition pdfs $p \left( \mathbf{x}_{k} | \tilde{\mathbf{x}}^a_{k-1,i} \right)$ as the importance density to generate the particles $\mathbf{x}^b_{k,i}$. One can also use other importance densities for this purpose, as discussed previously. For examples, see, \citet{Arulampalam2002,bocquet2010beyond,vanLeeuwen2010-nonlinear}.

\underline{Filtering step:} With an incoming observation $\mathbf{y}_{k}$, the particles remain unchanged, i.e., $\mathbf{x}^a_{k,i} = \mathbf{x}^b_{k,i}$, while the associated weights -- in light of the choice of the importance density $p \left( \mathbf{x}_{k} | \tilde{\mathbf{x}}^a_{k-1,i} \right)$ -- are updated according to Bayes' rule so that
\begin{linenomath*}
\begin{equation} \label{eq:weight_update_PF}
w_{k,i} = \frac{\tilde{w}_{k-1,i} \; p(\mathbf{y}_k | \mathbf{x}^b_{k,i})}{ \sum\limits_{i=1}^N \tilde{w}_{k-1,i} \; p(\mathbf{y}_k | \mathbf{x}^b_{k,i})}  \, ,
\end{equation}
\end{linenomath*}
where $p(\mathbf{y}_k | \mathbf{x}^b_{k,i})$ is the probability that $\mathbf{y}_k$ happens to be the observation with respect to $\mathbf{x}^b_{k,i}$.

\underline{Re-sampling step:} To overcome the problem of weight collapse, it is customary to introduce a re-sampling step to the PF when the effective sample size is below a certain threshold, or alternatively, when the difference between the weight ``entropy'' and that with the uniform weight exceeds a certain threshold (see Appendix \ref{sec:algorithm}). Many implementations of the PF distinguish each other mainly in their re-sampling strategies. There is a rich literature in this respect. Readers are referred to, for example, \cite{Arulampalam2002,VanLeeuwen2009} and the references therein on this issue. Here we only consider a re-sampling strategy based on the kernel density estimation (KDE), which leads to the so-called regularized particle filter (RPF) \cite[ch. 12]{Doucet2001-sequential}. In the RPF, one applies KDE to estimate a continuous pdf of the model state based on the particles $\mathbf{x}^a_{k,i}$ and their associated weights $w_{k,i}$, and then uses this pdf to draw $N$ new particles $\tilde{\mathbf{x}}^a_{k,i}$ ($i = 1, 2, \dotsb,N$), which are then assigned the identical weight $1/N$ after re-sampling. More details of the RPF implemented in this work are provided in Appendix \ref{sec:algorithm}.

\section{Residual nudging and its implementation in the particle filter}\label{sec:RPF_RN}

\subsection{Residual nudging} \label{subsec:RN}

For ease of discussion, we first define some notations. The weighted sample mean $\hat{\mathbf{x}}^a_k$ of the (analysis) particles $\mathbf{x}^a_{k,i}$ is given by
\begin{linenomath*}
\begin{equation}
\hat{\mathbf{x}}^a_k = \sum\limits_{i=1}^{N} w_{k,i} \mathbf{x}^a_{k,i} \, ,
\end{equation}
\end{linenomath*}
and the corresponding residual is $\hat{\mathbf{r}}_k^a \equiv \mathcal{H}_k (\hat{\mathbf{x}}^a_k) - \mathbf{y}_k^o$ with respect to the observation $\mathbf{y}_k^o = \mathcal{H}_k (\mathbf{x}^{tr}_k) + \tilde{\varepsilon}_k^o$ at instant $k$, where $\mathbf{x}^{tr}_k$ is the corresponding truth, and $\tilde{\varepsilon}_k^o$ a realization of the observation error. Define the weighted $\ell_2$-norm of a vector $\mathbf{z}$ as
\begin{linenomath*}
\begin{equation} \label{eq:2_norm}
\Vert \mathbf{z} \Vert_{\mathbf{W}} \equiv \sqrt{\mathbf{z}^T (\mathbf{W})^{-1} \mathbf{z}} \, ,
\end{equation}
\end{linenomath*}
where the normalization (or weight) matrix $\mathbf{W}$ is symmetric and positive definite. Throughout this work, $\mathbf{W}$ is chosen to be the covariance matrix $\mathbf{R}_k$, although there certainly exist other possibilities (also see the discussion in Section \ref{sec:RN discussion}).

Under the above setting, and by the triangle inequality, one has
\begin{linenomath*}
\begin{equation} \label{eq:residual_norm}
\Vert \hat{\mathbf{r}}_k^a \Vert_{\mathbf{R}_k} \leq \Vert \mathcal{H}_k (\hat{\mathbf{x}}^a_k) - \mathcal{H}_k (\mathbf{x}_k^{tr}) \Vert_{\mathbf{R}_k} + \Vert \tilde{\varepsilon}_k^o \Vert_{\mathbf{R}_k}  \, .
\end{equation}
\end{linenomath*}
For a reasonably good estimate $\hat{\mathbf{x}}^a_k$, we expect that in general $\Vert \mathcal{H}_k (\hat{\mathbf{x}}^a_k) - \mathcal{H}_k (\mathbf{x}_k^{tr}) \Vert_{\mathbf{R}_k}$ should not substantially exceed {\color{black}{the observation noise term $\Vert \tilde{\varepsilon}_k^o \Vert_{\mathbf{R}_k}$, which, in a certain sense, is connected to the number of independent observations (see the discussion below).}} On the other hand, we have $ (\mathbb{E} \Vert \tilde{\varepsilon}_k^o \Vert_{\mathbf{R}_k})^2 \leq \mathbb{E} \Vert \tilde{\varepsilon}_k^o \Vert_{\mathbf{R}_k}^2 = \text{trace} (\mathbb{E} (\tilde{\varepsilon}_k^o (\tilde{\varepsilon}_k^o)^T) \mathbf{R}_k^{-1} ) = \text{trace} (\mathbf{R}_k \mathbf{R}_k^{-1}) = p$, thus the expectation $\mathbb{E} \Vert \tilde{\varepsilon}_k^o \Vert_{\mathbf{R}_k}$ of the weighted $\ell_2$-norm of the observation noise $\tilde{\varepsilon}_k^o$ is (at most) in the order of $\sqrt{p}$. By requiring that a reasonably good estimate have $\Vert \mathcal{H}_k (\hat{\mathbf{x}}^a_k) - \mathcal{H}_k (\mathbf{x}_k^{tr}) \Vert_{\mathbf{R}_k}$ also in the order of $\sqrt{p}$ (or even less), one comes to that $\Vert \hat{\mathbf{r}}_k^a \Vert_{\mathbf{R}_k}$ should be upper bounded by $\beta \sqrt{p}$, where $\beta$ is a pre-chosen real positive scalar, called the noise level coefficient hereafter. The choice of $\beta$ will be further discussed later.

We introduce residual nudging to the PF after the filtering step, and before the re-sampling step. The objective in residual nudging is the following. We accept $\hat{\mathbf{x}}^a_k$ as a reasonable estimate if its residual norm $\Vert \hat{\mathbf{r}}_k^a \Vert_{\mathbf{R}_k}$ is no larger than the pre-specified value $\beta \sqrt{p}$. Otherwise, we consider $\hat{\mathbf{x}}^a_k$ a poor estimate, and thus look for a replacement, say, $\breve{\mathbf{x}}^a_k$, based on the original estimate $\hat{\mathbf{x}}^a_k$ and the observation $\mathbf{y}_k^o$, so that the new residual norm of $\breve{\mathbf{x}}^a_k$ is no larger than $\beta \sqrt{p}$.

In what follows we assume that the observation operator $\mathcal{H}_k$ is a linear operator (e.g., a matrix), denoted by $\mathbf{H}_k$ hereafter. For nonlinear observation operators, the procedures in residual nudging become more complicated. One may, for instance, linearize $\mathcal{H}_k$ locally as in the extended Kalman filter, or adopt a numerical optimization algorithm to find a replacement estimate. Investigations of these possible strategies will be considered in future work.

In case of linear observations, the objective in residual nudging can be achieved as follows. First of all, we construct a potentially new estimate $\breve{\mathbf{x}}^a_k$ by letting
\begin{linenomath*}
\begin{equation} \label{eq:replacement}
\breve{\mathbf{x}}^a_k = c_k  \, \hat{\mathbf{x}}^a_k + (1-c_k) \, \mathbf{x}^{o}_k \, ,
\end{equation}
\end{linenomath*}
where $c_k \in [0,1]$ is the fraction coefficient that will be calculated later, and
$\mathbf{x}^{o}_k$, termed observation inversion in this work, is a solution of the equation
\begin{linenomath*}
\begin{equation} \label{eq:obs_equ}
\mathbf{H}_k \mathbf{x}_k = \mathbf{y}^{o}_k \, .
\end{equation}
\end{linenomath*}
Under the assumption that the observation dimension $p$ is no larger than the system state dimension $n$, Eq. (\ref{eq:obs_equ}) may be an under-determined problem, i.e., the solutions of Eq. (\ref{eq:obs_equ}) may not be unique. Numerically, if $\mathbf{H}_k$ is of moderate dimension, one may choose to directly compute a pseudo-inverse of $\mathbf{H}_k$ \citep{Luo2012-residual}. On the other hand, if the dimension of $\mathbf{H}_k$ is large and it is inconvenient to compute a pseudo-inverse in a straightforward way, then there are a few alternative ways to find an observation inversion $\mathbf{x}^{o}_k$. One is to conduct a QR decomposition on $\mathbf{H}_k^T$ \citep{Luo2012-residual}; another is to directly apply an iterative optimization algorithm (e.g., conjugate gradient) to the linear equation Eq. (\ref{eq:obs_equ}) \citep{Engl2000-regularization,Nocedal-numerical}; and the third is to construct a merit function \citep{Nocedal-numerical}, which recasts the problem of solving a linear equation as a least squares problem, as described below.

To construct the merit function, we follow the custom in inverse problems (see, for example, \citealp{Engl2000-regularization}) and give preference to the solutions with relatively small magnitudes. Therefore we recast the problem of solving Eq. (\ref{eq:obs_equ}) as a weighted least squares problem, in the form of
\begin{linenomath*}
\begin{equation} \label{eq:ls_problem}
\underset{\mathbf{x}}{\text{\normalsize min}} \; \Vert \mathbf{H}_k \mathbf{x}_k - \mathbf{y}^{o}_k \Vert_{\mathbf{R}_k}^2 + \dfrac{1}{\alpha} \Vert \mathbf{x}_k \Vert_{\mathbf{\Omega}_k}^2 \, .
\end{equation}
\end{linenomath*}
The second term in (\ref{eq:ls_problem}) represents a regularization term that sorts out a preferred solution from the many possible ones. There $\alpha$ is a non-negative scalar and $\mathbf{\Omega}_k$ is the weight matrix associated with $\mathbf{x}_k$.

The specific choice of the regularization term in (\ref{eq:ls_problem}) is recommended to use only for the situations in which one does not have further ``prior knowledge'' (e.g., physical constraints like variable bounds and/or dynamical balance) of the model state. In this specific context, the ``prior knowledge'' does not include the information from the available particles themselves, since it is already represented by the original estimate $\hat{\mathbf{x}}^a_k$ in Eq. (\ref{eq:replacement}). With this argument, it is clear that the formulated least squares problem (\ref{eq:ls_problem}) only represents one - but by no means the best - possible choice in finding an observation inversion. If one does have certain ``prior knowledge'' of the model state, then it would be more appropriate to re-formulate the least squares problem to better reflect the availability of these extra information sources, e.g., in the form of a constrained optimization problem \citep{Nocedal-numerical}, {\color{black}{or by constructing a special weight matrix $\mathbf{\Omega}_k$ that enhances the model balance of the state estimate}}. In general the formation of such a regularization problem may be case-dependent, and is thus not pursued in this study.

In general, the least squares problem (\ref{eq:ls_problem}) can be solved in the framework of 3D-Var \citep{vanLeeuwen2010-nonlinear}. Specifically, when the observation operator is linear, an explicit solution can be obtained as follows
\begin{linenomath*}
\begin{equation} \label{eq:pseudo_inversion}
\mathbf{x}^{o}_k = ( \alpha \mathbf{\Omega}_k ) \mathbf{H}_k^T ( \mathbf{H}_k ( \alpha \mathbf{\Omega}_k ) \mathbf{H}_k^T + \mathbf{R}_k)^{-1} \, \mathbf{y}^{o}_k \, .
\end{equation}
\end{linenomath*}
If one treats $\alpha$ as a covariance inflation factor, then Eq. (\ref{eq:pseudo_inversion}) corresponds to a Kalman update scheme with the background mean and inflated covariance being $0$ and $\alpha \mathbf{\Omega}_k$, respectively (and the zero background mean is consistent with our solution preference in solving the under-determined linear equation). This point of view motivates us to take $\mathbf{\Omega}_k$ as the background sample covariance $\mathbf{P}_k^b$ of the particles (with equal weights). However, in light of the fact that the least squares problem (\ref{eq:ls_problem}) requires $\mathbf{\Omega}_k$ to be of full rank, we follow the idea in the hybrid EnKF \citep{Hamill-hybrid} and choose $\mathbf{\Omega}_k$ to be a hybrid of $\mathbf{P}_k^b$ and a background covariance $\mathbf{B}$ which can be obtained from, for example, a long-term run of the dynamical model (see the descriptions in the experiments later). More concretely, we let
\begin{linenomath*}
\begin{equation} \label{eq:hybrid_cov}
\mathbf{\Omega}_k = 0.5 \, \mathbf{P}_k^b + 0.5 \, \mathbf{B} \, ,
\end{equation}
\end{linenomath*}
throughout this work. On the other hand, as an approximate solution to Eq. (\ref{eq:obs_equ}), we are interested in obtaining an observation inversion $\mathbf{x}^{o}_k$ with a relatively small residual norm $ \Vert \mathbf{r}_k^o \Vert_{\mathbf{R}_k} \equiv \Vert \mathbf{H}_k \mathbf{x}^{o}_k - \mathbf{y}^{o}_k \Vert_{\mathbf{R}_k}$. To this end, we choose a relatively (very) large value for $\alpha$. Specifically, we let
\begin{linenomath*}
\begin{equation} \label{eq:darn_alpha}
\alpha = 10^{10} \times  \text{trace} (\mathbf{R}_k) / \text{trace} (\mathbf{H}_k \mathbf{\Omega}_k \mathbf{H}_k^T)
\end{equation}
\end{linenomath*}
in this work.

Next we need to choose a proper fraction coefficient $c_k$ so that the residual norm with respect to the new estimate $\breve{\mathbf{x}}^a_k$ is no larger than $\beta \sqrt{p}$. We consider two possibilities: (a) the original residual norm $ \Vert \hat{\mathbf{r}}^a_k \Vert_{\mathbf{R}_k}  \leq \beta \sqrt{p}$. In this case, we do not introduce any change to the original estimate $\hat{\mathbf{x}}^a_k$, and choose $c_k = 1$ in Eq.~(\ref{eq:replacement}); and (b) the original residual norm $ \Vert \hat{\mathbf{r}}^a_k \Vert_{\mathbf{R}_k}> \beta \sqrt{p}$. In this case, by applying $\mathbf{H}_k $ to both sides of Eq.~(\ref{eq:replacement}), we have the new residual $\breve{\mathbf{r}}^a_k = \mathbf{H}_k \breve{\mathbf{x}}^a_k - \mathbf{y}^{o}_k = c_k \, \hat{\mathbf{r}}^a_k + (1 - c_k) \, \mathbf{r}_k^o$. By applying the triangle inequality again to the new residual norm, it can be shown that a sufficient condition to guarantee $ \Vert \breve{\mathbf{r}}^a_k \Vert_{\mathbf{R}_k}  \leq \beta \sqrt{p}$ is to choose $ c_k \leq (\beta \sqrt{p} - \Vert \mathbf{r}_k^o \Vert_{\mathbf{R}_k}) / ( \Vert \hat{\mathbf{r}}^a_k \Vert_{\mathbf{R}_k} - \Vert \mathbf{r}_k^o \Vert_{\mathbf{R}_k})$. Throughout this work we choose $c_k = (\beta \sqrt{p} - \Vert \mathbf{r}_k^o \Vert_{\mathbf{R}_k})/(\Vert \hat{\mathbf{r}}^a_k \Vert_{\mathbf{R}_k} - \Vert \mathbf{r}_k^o \Vert_{\mathbf{R}_k})$.
One may also take smaller values for $c_k$, whose effect is then equivalent to taking smaller $\beta$ values. Combining the above two possibilities, one may re-write the choice of $c_k$ in a more compact form, i.e.,
\begin{linenomath*}
\begin{equation} \label{eq:fraction_coefficient}
c_k = \min \left( 1, \dfrac{\beta \sqrt{p} - \Vert \mathbf{r}_k^o \Vert_{\mathbf{R}_k}}{\Vert \hat{\mathbf{r}}^a_k \Vert_{\mathbf{R}_k} - \Vert \mathbf{r}_k^o \Vert_{\mathbf{R}_k}} \right) \, .
\end{equation}
\end{linenomath*}

After obtaining the new analysis mean $\breve{\mathbf{x}}^a_k$ through the above procedures, in general one may need to find a new set of particles
 $\breve{\mathbf{x}}_{k,i}^a$ and the associated weights $\breve{w}_{k,i}$, so that $\breve{\mathbf{x}}^a_k = \sum\limits_{i=1}^N \breve{w}_{k,i} \, \breve{\mathbf{x}}_{k,i}^a$. In doing so, it is equivalent to making certain modifications to the empirical posterior pdf in Eq. (\ref{BRR:update}) of the PF, so that the modified empirical posterior pdf may not be equivalent to the original one any more. As we have discussed in Section \ref{sec:introduction}, in certain circumstances (e.g., when with a relatively small sample size) the original empirical posterior pdf of the PF may be a poor approximation to the truth. In such circumstances it would appear reasonable to introduce a certain correction to the original empirical posterior pdf, instead of using it for subsequent procedures. In this regard, residual nudging may be considered as a technique that, when necessary, provides a correction to the mean of the empirical pdf. One might also come up with other higher-order moments correction schemes.

In the equation $\breve{\mathbf{x}}^a_k = \sum\limits_{i=1}^N \breve{w}_{k,i} \, \breve{\mathbf{x}}_{k,i}^a$, the weights $\breve{w}_{k,i}$ may depend on $\breve{\mathbf{x}}_{k,i}^a$, for instance, by letting $\breve{w}_{k,i} \propto p(\mathbf{y}_k^o | \breve{\mathbf{x}}_{k,i}^a)$. Therefore, in general one needs to solve $n$ (scalar, under-determined) nonlinear equations with $N \times n$ unknowns whenever residual nudging is conducted, which may appear complicated and expensive in high dimensional systems. Here we adopt a heuristic, yet simple strategy. We let the weights associated with the new particles be $\breve{w}_{k,i} = w_{k,i}$. In addition, we preserve the deviations to the analysis mean so that $\breve{\mathbf{x}}_{k,i}^a - \breve{\mathbf{x}}_{k}^a = \mathbf{x}_{k,i}^a - \hat{\mathbf{x}}_{k}^a$. Under this choice, we have $\breve{\mathbf{x}}_{k,i}^a  = \breve{\mathbf{x}}_{k}^a + (\mathbf{x}_{k,i}^a - \hat{\mathbf{x}}_{k}^a)$,  $\sum\limits_{i=1}^N \breve{w}_{k,i} \, \breve{\mathbf{x}}_{k,i}^a = \breve{\mathbf{x}}^a_k$, and the weighted sample covariance $\sum\limits_{i=1}^N \breve{w}_{k,i} \, (\breve{\mathbf{x}}_{k,i}^a - \breve{\mathbf{x}}^a_k)(\breve{\mathbf{x}}_{k,i}^a - \breve{\mathbf{x}}^a_k)^T$ of $\breve{\mathbf{x}}_{k,i}^a$ is equal to that of $\mathbf{x}_{k,i}^a$. By Eq.~(\ref{eq:replacement}) we have
\begin{linenomath*}
\begin{equation} \label{eq:particles_nudging}
\breve{\mathbf{x}}_{k,i}^a  = \mathbf{x}_{k,i}^a  + (\breve{\mathbf{x}}_{k}^a - \hat{\mathbf{x}}_{k}^a) = \mathbf{x}_{k,i}^a  + (1-c_k) (\mathbf{x}^{o}_k -  \hat{\mathbf{x}}_{k}^a) \, ,
\end{equation}
\end{linenomath*}
which implies that the new set of particles is simply a spatial translation of the original one.

Let $\mathbf{y}_{k,i}^a = \mathbf{H}_k \mathbf{x}_{k,i}^a$, $\breve{\mathbf{y}}_{k,i}^a = \mathbf{H}_k \breve{\mathbf{x}}_{k,i}^a$ and $\hat{\mathbf{y}}_{k}^a = \mathbf{H}_k \hat{\mathbf{x}}_{k}^a$ be the projections of $\mathbf{x}_{k,i}^a$, $\breve{\mathbf{x}}_{k,i}^a$ and $\hat{\mathbf{x}}_{k}^a$ onto the observation space, respectively, then $\breve{\mathbf{y}}_{k,i}^a  = \mathbf{y}_{k,i}^a  + (1-c_k) (\mathbf{y}^{o}_k -  \hat{\mathbf{y}}_{k}^a)$. Therefore, when $c_k<1$ such that $1-c_k > 0$ (i.e., when the residual norm $\Vert \hat{\mathbf{r}}_k^a \Vert_{\mathbf{R}_k}$ exceeds the threshold $\beta \sqrt{p}$), residual nudging tends to move the particle projections toward the observation $\mathbf{y}^{o}_k$ at the same assimilation cycle, with identical length and direction of movement in the observation space. Through some experiments later, we show that the nudging strategy Eq. (\ref{eq:particles_nudging}) improves the performance of the PF, in terms of filter accuracy and/or stability against divergence. We note that it is also possible for one to introduce nudging terms to the particles through other strategies. For instance, \citet{vanLeeuwen2010-nonlinear,ades2012exploration} introduce nudging terms to the particles and use them as the samples of the chosen importance density. Comparison and combination of different possible nudging strategies will be considered elsewhere.

After residual nudging is done, the re-sampling step (if any) is then conducted with respect to the new particles $\breve{\mathbf{x}}_{k,i}^a$ and the associated weights $\breve{w}_{k,i}$. The subsequent prediction and filtering steps for the next data assimilation cycle are the same as in the normal particle filter, followed by another residual nudging step if necessary, and so on.

\subsection{Discussion} \label{sec:RN discussion}

It is worth to discuss what the differences may be if one replaces the covariance matrix $\mathbf{R}_k$ by a general symmetric and positive definite matrix $\mathbf{W}_k$ in (\ref{eq:residual_norm}). In that case, one also has an inequality reading $ (\mathbb{E} \Vert \tilde{\varepsilon}_k^o \Vert_{\mathbf{W}_k})^2 \leq \mathbb{E} \Vert \tilde{\varepsilon}_k^o \Vert_{\mathbf{W}_k}^2 = \text{trace} (\mathbb{E} (\tilde{\varepsilon}_k^o (\tilde{\varepsilon}_k^o)^T) \mathbf{W}_k^{-1} ) = \text{trace} (\mathbf{R}_k \mathbf{W}_k^{-1})$. As a result, Eq. (\ref{eq:fraction_coefficient}) becomes
\begin{linenomath*}
\[
c_k = \min \left(1, \dfrac{\beta \sqrt{\text{trace} (\mathbf{R}_k \mathbf{W}_k^{-1})} - \Vert \mathbf{r}_k^o \Vert_{\mathbf{W}_k}}{\Vert \hat{\mathbf{r}}_k^a \Vert_{\mathbf{W}_k} - \Vert \mathbf{r}_k^o \Vert_{\mathbf{W}_k}} \right),
\]
\end{linenomath*}
while the subsequent equations, e.g., Eqs. (\ref{eq:replacement}) and (\ref{eq:particles_nudging}), remain unchanged. Therefore, for a given $\beta$, the choice of $\mathbf{W}_k$ only affects the value of the fraction coefficient $c_k$, which in effect is equivalent to varying the noise level coefficient $\beta$ given a fixed normalization matrix, say, $\mathbf{R}_k$. Therefore, in this work we do not investigate the effects of different normalization matrices $\mathbf{W}_k$. Instead, we examine the impact of $\beta$ on residual nudging in some experiments later.

Once a noise level coefficient $\beta$ is chosen, we keep it constant over the whole assimilation time window. However, the corresponding fraction coefficient $c_k$ in Eq.~(\ref{eq:fraction_coefficient}) may vary from time to time, so that the new analysis $\breve{\mathbf{x}}^a_k$ in Eq. (\ref{eq:replacement}) is an adaptive combination of the original analysis $\hat{\mathbf{x}}^a_k$ and the observation inversion $\mathbf{x}_k^o$. Roughly speaking, the choice of $\beta$ reflects the relative confidence of the filter designer in $\hat{\mathbf{x}}^a_k$ and $\mathbf{x}_k^o$. A small $\beta$ means that the filter designer tends to rely heavily on $\mathbf{x}_k^o$, while a large $\beta$ implies that the filter designer wants $\hat{\mathbf{x}}^a_k$ to be dominant. These can be seen from Eqs.~(\ref{eq:fraction_coefficient}) and (\ref{eq:replacement}). Because $c_k \in [0,1]$,  the new analysis $\breve{\mathbf{x}}^a_k$ in Eq.~(\ref{eq:replacement}) is a convex combination of  $\hat{\mathbf{x}}^a_k$ and $\mathbf{x}^{o}_k$, i.e., an estimate somewhere in-between the original estimate $\hat{\mathbf{x}}^a_k$ and the observation inversion $\mathbf{x}^{o}_k$, depending on the value of $c_k$. If one chooses a large value for $\beta$, or, if for a fixed $\beta$ the original residual norm $\hat{\mathbf{r}}_k^a$ is sufficiently small, then the fraction coefficient $c_k \rightarrow 1$ according to Eq.~(\ref{eq:fraction_coefficient}), thus $\breve{\mathbf{x}}^a_k \rightarrow \hat{\mathbf{x}}^a_k$ according to Eq.~(\ref{eq:replacement}). Therefore $\breve{\mathbf{x}}^a_k$ will be a good estimate if $\hat{\mathbf{x}}^a_k$ is so (as will be further discussed later), but may not be able to achieve a good estimation accuracy when $\hat{\mathbf{x}}^a_k$ itself is poor. On the other hand, if one chooses a very small value for $\beta$, or, if for a fixed $\beta$ the original residual norm $\hat{\mathbf{r}}_k^a \rightarrow +\infty$ (e.g., with filter divergence), then $c_k \rightarrow 0$, $\breve{\mathbf{x}}^a_k \rightarrow \mathbf{x}^{o}_k$, and $ \Vert \breve{\mathbf{r}}^a_k \Vert_{\mathbf{R}_k} \rightarrow 0$. In this case, the estimate $\breve{\mathbf{x}}^a_k$ is calculated mainly based on the information content of the observation $\mathbf{y}^{o}_k$, and may result in a relatively poor accuracy due to the existence of the observation noise $\varepsilon_k^o$ in Eq.~(\ref{ps_observation_system}), together with the ignorance of the prior information about the dynamical model. However, using $\mathbf{x}^{o}_k$ as the estimate is often a relatively safe (although conservative) strategy, in that for a given observation $\mathbf{y}^{o}_k$, $\mathbf{x}^{o}_k$ tends to be less sensitive to the model error and the sample size (hence the effect of weight collapse). 

Our main objective in this study is to present residual nudging as a safeguard strategy for the PF in case that it does not perform well in certain circumstances, due to, for instance, the small sample size. However, it may still be of interest to gain some insights on the asymptotic behaviour of the PF with residual nudging. For instance, what happens if one introduces residual nudging to a PF which, with infinitely many particles, converges to the optimal solution. In such cases, the PF with residual nudging can have the same optimal solution as the normal PF, provided that $\beta$ is sufficiently large. This can be achieved by making the fraction coefficients $c_k = 1$ for all $k$, such that by Eqs. (\ref{eq:replacement}) and (\ref{eq:particles_nudging}) residual nudging will have no effect on the original PF solution. To guarantee $c_k = 1$ for all $k$, a sufficient condition is to make $\beta \sqrt{p}/\Vert \hat{\mathbf{r}}^{a}_k \Vert_{\mathbf{R}_k}$ in Eq. (\ref{eq:fraction_coefficient}) no less than $1$, which implies that $\beta \geq \max \limits_{k} \Vert \hat{\mathbf{r}}^{a}_k \Vert_{\mathbf{R}_k}/\sqrt{p}$. In this aspect, a more convenient strategy would be to make $\beta$ adaptive with time, rather than pre-set it over the whole assimilation window. Given our main objective in this study, though, we do not consider the adaptive choice of $\beta$.

\section{Numerical results in a linear scalar model} \label{sec:example_scalar}

First we investigate the performance of the RPF and RPF-RN in a scalar, first order autoregressive (AR1) model driven by Gaussian white noise. The motivation in conducting this experiment is the following. Due to the low dimensionality of the model, the estimate of the normal PF would approach the optimal one with a reasonably small sample size (in terms of computational cost). This provides a computationally convenient platform to investigate the behaviour of the PF with residual nudging when the normal PF is performing well.

In the experiment the scalar AR1 model is given by
\begin{linenomath*}
\begin{equation} \label{eq:ar1}
x_{k+1} = 0.9 \, x_k + \varepsilon_k^m \, ,
\end{equation}
\end{linenomath*}
where $\varepsilon_k^m$ represents the dynamical noise, which follows the Gaussian distribution with zero mean and variance 1, and is thus denoted by $\varepsilon_k^m \sim N(\varepsilon_k^m:0,1)$. The observation model is described by
\begin{linenomath*}
\begin{equation} \label{eq:ar1_obs}
y_{k} = x_k + \varepsilon_k^o \, ,
\end{equation}
\end{linenomath*}
where $\varepsilon_k^o \sim N(\varepsilon_k^o:0,1)$ is the observation noise, and is uncorrelated with $\varepsilon_k^m$.

The filters adopted in this work are based on the regularized particle filter (RPF) (\citealp[ch. 12]{Doucet2001-sequential}, also see Appendix \ref{sec:algorithm}). We compare the performance of the normal RPF and that of the RPF equipped with residual nudging (RPF-RN). In the experiment, we integrate the AR1 model forward for $10,000$ steps (integration steps hereafter), with the initial value $x_0$ randomly drawn from the Gaussian distribution $N(0,1)$, and the associated initial prior variance being 1. The true states (truth) $\{ x_k \}_{i=0}^{9999}$ are obtained by drawing samples of dynamical noise from the distribution $N(0,1)$, and adding them to $x_k$ to obtain $x_{k+1}$ at the next integration step, and so on. The synthetic observations $y^o_k$ are obtained by adding to the model state $x_{k}$ samples of observation noise from the distribution $N(0,1)$, and are assimilated into the AR1 model every 4 integration steps. The RPF has an ensemble of $1000$ particles, which are initialized by drawing samples from $N(0,1)$. Except for the presence of residual nudging and the related procedures, the RPF-RN has the same experiment settings as the normal RPF. In the RPF-RN, the background covariance $\mathbf{B}$ in Eq. (\ref{eq:hybrid_cov}) is obtained by integrating the AR1 model forward for $100,000$ steps and taking $\mathbf{B}$ as the temporal covariance of the corresponding trajectory. The noise level coefficient $\beta$ in the RPF-RN is taken from the set $\{0.02, 0.2, 1, 2:2:20\}$, where the notation $v_i:\delta v: v_f$ means a set of values that grows from the initial value $v_i$ to the final one $v_f$, with an even increment $\delta v$ each time. To reduce statistical fluctuations, we repeat the experiment $20$ times, each time with randomly drawn $x_0$, $\varepsilon_k^m$, $\varepsilon_k^o$ and the initial particles of the filters.

We use the average root mean squared error (average RMSE) to measure the accuracy of a filter estimate. For an $n$-dimensional system, the RMSE $e_k$ of an estimate $\hat{\mathbf{x}}_k = [\hat{x}_{k,1}, \dotsb, \hat{x}_{k,n}]^T$ with respect to the true state vector $\mathbf{x}_k^{tr} = [x_{k,1}^{tr}, \dotsb, x_{k,n}^{tr}]^T$ at time instant $k$ is defined as
\begin{linenomath*}
\begin{equation} \label{eq:def_of_rmse}
e_k = \Vert \hat{\mathbf{x}}_k - \mathbf{x}_k^{tr} \Vert_{\mathbf{I}_n} /\sqrt{n} \, ,
\end{equation}
\end{linenomath*}
{\color{black}{where $\mathbf{I}_n$ denotes the $n$-dimensional identity matrix}}.The average RMSE $\hat{e}_k$ at time instant $k$ over $M$ repetitions of the same experiment is thus defined as $\hat{e}_k =\sum_{j=1}^{M} e_k^j/M$ ($M=20$ in our setting), where $e_k^j$ denotes the RMSE at time instant $k$ in the $j$th repetition of the experiment. We also define the time mean RMSE $\hat{e}$ as the average of $\hat{e}_k$ over the assimilation time window with $S$ integration steps, i.e., $\hat{e} = \sum_{i=0}^{S-1} \hat{e}_k/S$ ($S = 10000$ here). One may also adopt other metrics (e.g., a certain weighted norm as in Section \ref{sec:RPF_RN}), rather than the Euclidean norm in Eq. (\ref{eq:def_of_rmse}), as the performance measure. Since it is customary to use the Euclidean norm in the literature, we stick to this choice hereafter.

\begin{figure*} 
\vspace*{2mm}
\centering

\includegraphics[width=\textwidth]{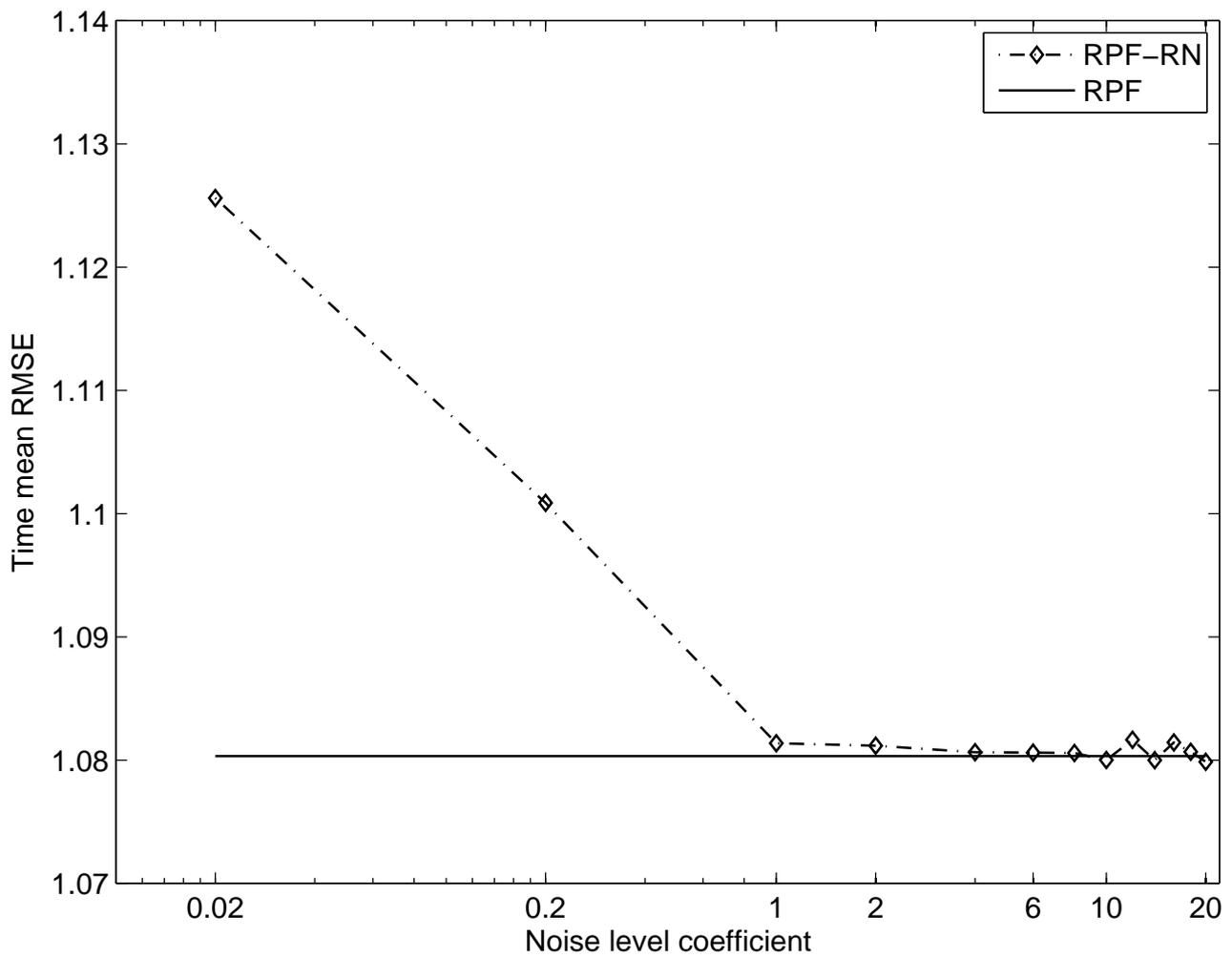}

\caption{\label{fig:AR1_RPF_RN_varying_beta_ana_rmse} Time mean RMSEs of the RPF and RPF-RN as functions of the noise level coefficient $\beta$.}
\end{figure*}

Fig. \ref{fig:AR1_RPF_RN_varying_beta_ana_rmse} shows the time mean RMSEs of the RPF and RPF-RN as functions of the noise level coefficient $\beta$. Because of the different orders of magnitudes of $\beta$ used in the experiment, the horizontal axis is plotted in the logarithmic scale. The time mean RMSE of the RPF is around $1.08$\footnote{For reference, the corresponding time mean RMSE of the Kalman filter is about $1.06$ \citep{Luo2012-residual}.}, independent of $\beta$, therefore the corresponding curve is a horizontal line. For the RPF-RN, its performance depends on $\beta$. Starting from $\beta = 0.02$, the time mean RMSE of the RPF-RN tends to decrease as $\beta$ grows, until $\beta$ reaches $10$. After that, there are some slight fluctuations as $\beta$ grows further. The behaviour of the RPF-RN is largely consistent with our discussion in Section \ref{sec:RN discussion}. Indeed, when $\beta$ is relatively small, the fraction coefficient $c_k$ in Eq. (\ref{eq:fraction_coefficient}) tends to be smaller, thus by Eq. (\ref{eq:replacement}) the observation inversion has a larger impact on the estimate of the RPF-RN, while the reasonably good original state estimate may be under-represented. As a result, the performance of the RPF-RN is relatively poor in comparison to the normal RPF. As $\beta$ increases, $c_k$ approaches $1$, hence the original state estimate becomes more influential, so that the performance of the RPF and RPF-RN becomes close to each other.

\begin{figure*} 
\vspace*{2mm}
\centering
\subfigure[Time series of the fraction coefficient of the RPF-RN at $\beta = 0.2$]{ \label{subfig:AR1_fraction_coefficient_vs_time_beta02}
\includegraphics[scale = 0.7]{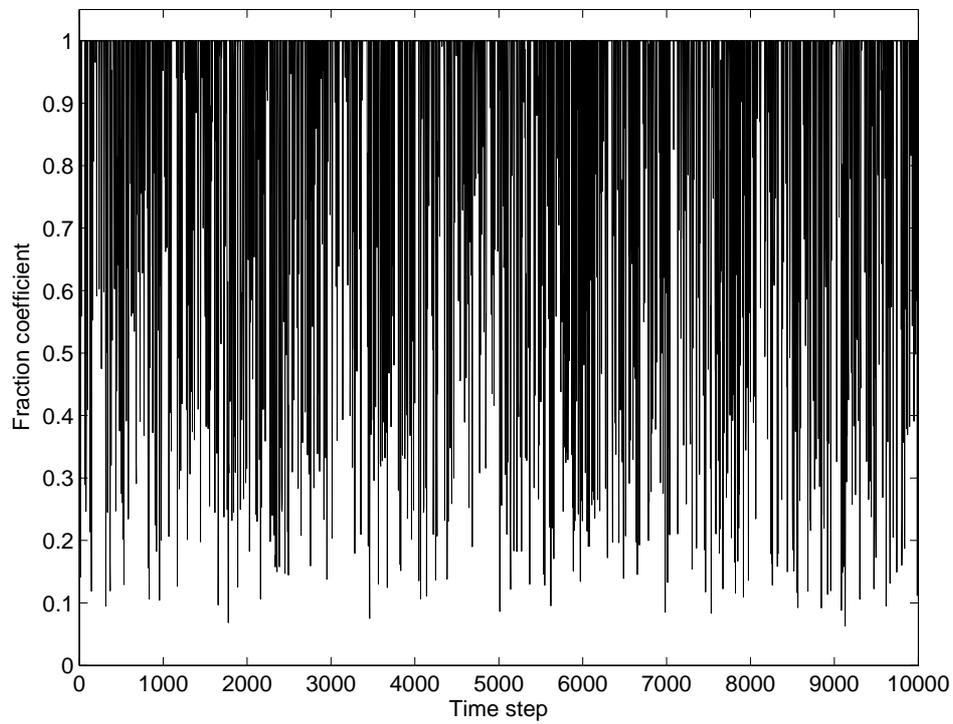}
}

\subfigure[Time series of the fraction coefficient of the RPF-RN at $\beta = 2$]{ \label{subfig:AR1_fraction_coefficient_vs_time_beta2}
\includegraphics[scale = 0.7]{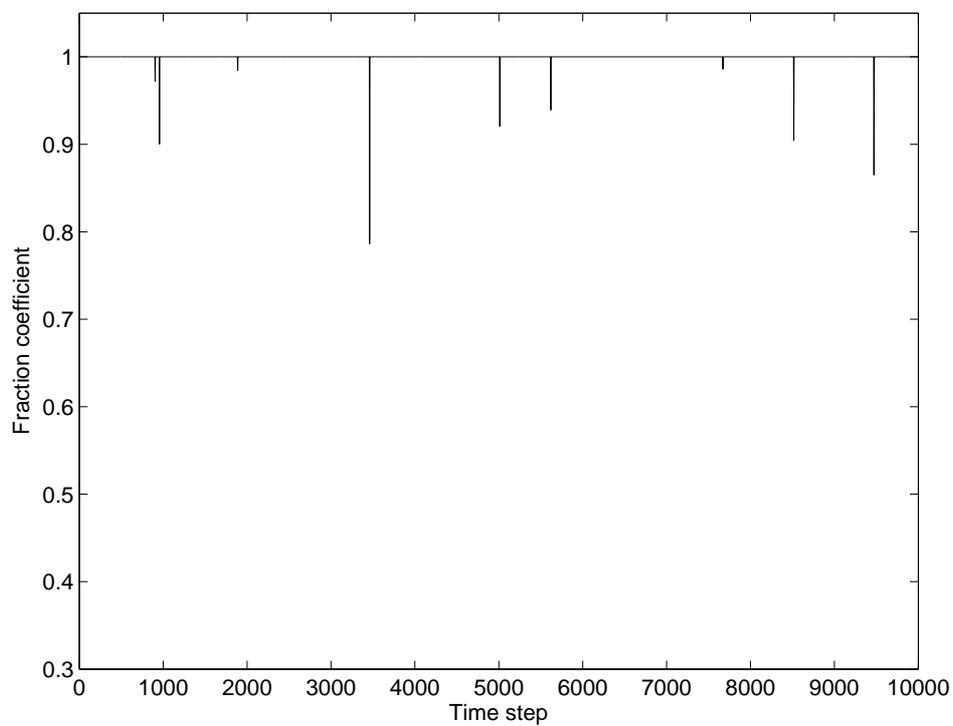}
}

\caption{\label{fig:AR1_fraction_coefficient_vs_time} Time series of the fraction coefficients of the RPF-RN. Panel (a): $\beta = 0.2$; Panel (b): $\beta = 2$. }
\end{figure*}

Fig. \ref{fig:AR1_fraction_coefficient_vs_time} depicts the time series of the fraction coefficients of the RPF-RN. At $\beta = 0.2$ (upper panel), there is a significant number of $c_k$ values that are relatively low, with $460$ out of $2500$ $c_k$ values being less than $0.5$, and the mean value $\hat{c}$ of $c_k$ being $0.8206$. In contrast, at $\beta = 2$ (lower panel), the fraction coefficient tends to be larger. Only $9$ out of $2500$ $c_k$ values are less than $1$, while the mean value $\hat{c}$ is $0.9997$. In both cases, though, the mean values $\hat{c}$ are quite close to $1$, meaning that $1 - \hat{c}$ are relatively small. Thus by Eqs. (\ref{eq:replacement}) and (\ref{eq:particles_nudging}), the particles in the RPF and those in the RPF-RN may not be significantly different. This might explain why even with a small value of $\beta$, say at $\beta = 0.2$, the time mean RMSE of the RPF-RN remains quite close to that of the RPF.

The above results suggest that it may not be very meaningful to introduce residual nudging to the PF when it already performs reasonably well. However, in many data assimilation practices, the dimensionality of the problems is often very high. Thus it may be prohibitively expensive to run a PF with a sufficiently large sample size in order for the filter to achieve good perform. On the other hand, the PF may perform poorly when running only with a finite, relatively small, sample size. Through the experiments below, we show that in cases that the PF does not perform well, equipping the PF with residual nudging may (substantially) enhance the filter performance, in terms of filter accuracy and/or stability against divergence.

\section{Numerical results in the 40-dimensional L95 model} \label{sec:example}

\subsection{Experiment settings} \label{sec:exp_setting}
We use the $40$-dimensional L95 model \citep{Lorenz-optimal} as the testbed. The governing equations of the L95 model are given by
\begin{linenomath*}
\begin{equation} \label{eq:LE98}
\frac{dx_i}{dt} = \left( x_{i+1} - x_{i-2} \right) x_{i-1} - x_i + F, \, i=1, \dotsb, 40.
\end{equation}
\end{linenomath*}
The quadratic terms simulate advection, the linear term represents internal dissipation, and $F$ acts as the external forcing term \citep{Lorenz-predictability}. Throughout this work, we choose $F = 8$ unless otherwise stated. For consistency, we define $x_{-1}=x_{39}$, $x_{0}=x_{40}$, and $x_{41}=x_{1}$ in Eq.~(\ref{eq:LE98}), and construct the state vector $\mathbf{x} \equiv [x_1,x_2,\dotsb,x_{40}]^T$.

We use the fourth-order Runge-Kutta method to integrate (and discretize) the system from time $0$ to $75$, with a constant integration step of $0.05$. To avoid the transition effect, we discard the trajectory between $0$ and $25$, and use the rest (with overall $1000$ integration steps) for data assimilation. The synthetic observation $\mathbf{y}_k$ is obtained by measuring (with observation noise) every $d$ elements of the state vector $\mathbf{x}_k = [x_{k,1},x_{k,2},\dotsb,x_{k,40}]^T$ at time instant $k$ ($k = 1, \dotsb, 1000$), i.e.,
\begin{linenomath*}
\begin{equation} \label{eq:L96_obs}
\mathbf{y}_k = \mathbf{H}^d \mathbf{x}_k + \varepsilon_k^o \, ,
\end{equation}
\end{linenomath*}
where $\mathbf{H}^d$ is a $(J+1) \times 40$ matrix such that $\mathbf{H}^d \mathbf{x}_k = [x_{k,1}, x_{k,1+d},\dotsb,x_{k,1+Jd}]^T$, with $J = \text{floor}(39/d)$ being the largest integer that is less than, or equal to, $39/d$, and $\varepsilon_k^o$ is the observation noise following the Gaussian distribution $N(\varepsilon_k^o:\mathbf{0},\mathbf{I}_{J+1})$, with $\mathbf{I}_{J+1}$ being the $(J+1)$-dimensional identity matrix. The elements $(\mathbf{H}^d)_{pq}$ of the matrix $\mathbf{H}^d$ can be determined as follows.
\begin{linenomath*}
\[
(\mathbf{H}^d)_{pq} = 1~\text{if}~q=(p-1)d+1\, , ~\text{otherwise}~(\mathbf{H}^d)_{pq} = 0 \, ,
\]
\end{linenomath*}
for $p = 1, \dotsb, (J+1), ~ q = 1,\dotsb, 40$. In all the experiments below the observations are made for every $4$ integration steps unless otherwise stated.

The filters in the experiments are configured as follows. To generate the initial particles, we run the L95 model from $0$ to $2500$ (overall $50000$ integration steps), and compute the temporal mean and covariance of the trajectory (the obtained temporal covariance is also used as the background covariance $\mathbf{B}$ in Eq. (\ref{eq:hybrid_cov})). We then assume that the initial state vectors follow a Gaussian distribution with the same mean and covariance, and draw a specified number of samples as the initial particles. {\color{black}{In many of the experiments, the sample sizes are relatively small so that the phenomenon of weight collapse is very severe.}} To mitigate this problem, we introduce a ``jittering'' procedure to the re-sampling step of the RPF following \citet{Gordon1993} (one may achieve a similar effect by increasing the bandwidth of the RPF). Concretely, after the re-sampling step of the RPF is finished, a random perturbation drawn from the Gaussian distribution $N(\mathbf{0},0.01 \times \mathbf{I}_{40})$ is added to each generated particle. Our experience shows that introducing ``jittering'' to the normal RPF improves the performance of the filter, especially in the case of small sample sizes. We note that the performance improvement of the RPF-RN over the normal RPF, as will be shown soon, does not depend on whether ``jittering'' is introduced or not. Performance improvement similar to what will be presented below was also observed when no ``jittering'' was introduced (results not reported).

To reduce statistical fluctuations, we repeat each experiment below for $20$ times, each time with randomly drawn initial state vectors of the L95 model, initial particles and observations. Except for the introduction of residual nudging, in all experiments the RPF and RPF-RN have identical configurations and experiment settings.

\subsection{Experiment results}

\subsubsection{Results with different observation operators}

Here we consider four different observation operators $\mathbf{H}^d$, with $d = 1, 2, 4, 8$, respectively. For convenience, we refer to them as the full, 1/2, 1/4 and 1/8 observation scenarios, respectively. The concrete configurations of the RPF and the RPF-RN are the following. In both filters the sample sizes are fixed to be $20$. In the RPF-RN, we let the noise level coefficient $\beta \in  \{0.02, 0.2, 1, 2:2:20\}$.

The time mean RMSEs (over $20$ repetitions) of the normal RPF are $4.8389$, $4.8963$, $4.8966$ and $4.9303$ in the full, 1/2, 1/4 and 1/8 observation scenarios, respectively. This shows that as the number of assimilated observations decreases, the time mean RMSE of the RPF becomes larger.

The time mean RMSE of the RPF-RN as a function of the noise level coefficient $\beta$ is shown in Fig. \ref{fig:DARN_RPF_rmse_vs_beta} (dash-dotted lines marked with diamonds), in which, for references, the corresponding time mean RMSEs of the normal RPF are also plotted as solid horizontal lines (since they do not depend on $\beta$). In the full observation case (upper left panel), when $\beta$ is small, say $\beta = 0.02$, the time mean RMSE is close to 1. This is expected, since in this case, $\beta \rightarrow 0$ implies that $c_k \rightarrow 0$ according to Eq.~(\ref{eq:fraction_coefficient}), and $\breve{\mathbf{x}}_k^a \rightarrow \mathbf{y}_k^o$ in the full observation scenario according to Eq.~(\ref{eq:replacement}), whose time mean RMSE should thus be equal to 1, due to the fact that the observation error covariance is $\mathbf{I}_{40}$, and that the L95 model has no dynamical noise (or very little due to ``jittering''). As $\beta$ increases, the time mean RMSE of the RPF-RN tends to decrease until $\beta$ reaches $6$, which achieves the minimum time mean RMSE $0.7789$, substantially lower than the corresponding value $4.8389$ in the RPF. Beyond that, continuing increasing $\beta$ would deteriorate the filter performance instead. Overall, the estimate $\hat{\mathbf{x}}_k^a$ of the RPF appears less informative than the observation inversion $\mathbf{x}^o_k$ in the sense that $\hat{\mathbf{x}}_k^a$ yields larger time mean RMSE than $\mathbf{x}^o_k$ (the estimate of the RPF-RN at a $\beta \rightarrow 0$).

The time mean RMSEs of the RPF-RN in the 1/2,1/4 and 1/8 observation scenarios exhibit behaviours similar to that in the full observation scenario. They all tend to decrease as $\beta$ grows from $0.02$. However, for the 1/2 and 1/4 observation scenarios, they achieve their minimum time mean RMSEs around $\beta = 20$, while for the 1/8 observation scenario it is around $\beta = 16$. Clearly, in all these three scenarios, the time mean RMSEs with $\beta \rightarrow 0$ are still (much) lower than those of the normal RPF, showing again that, in this specific context, the estimate of the RPF (with $20$ particles) is less informative than the observation inversion. In addition, there are even larger gaps between the minimum time mean RMSEs of the RPF-RN and the corresponding RMSEs of the normal RPF. This shows that a proper choice of $\beta$ can lead to further performance improvement (in terms of filter accuracy) of the RPF-RN, in contrast to just choosing the observation inversion as the estimate.

The upper panel of Fig. \ref{fig:RPFL96_corr_normalRPF_fraction_coefficient} shows a sample time series of the RMSEs of the RPF and RPF-RN ($\beta = 2$) in the 1/2 observation scenario, and the lower panel indicates the time series of the corresponding fraction coefficient of the RPF-RN. The RMSEs of the RPF-RN are lower than those of the RPF for a large proportion of the assimilation time window, and the corresponding fraction coefficients of the RPF-RN tend to be relatively small. Only $13$ out of the $250$ coefficients are larger than $0.1$, and the mean value of these $250$ coefficients is $0.0552$, indicating that in Eq. (\ref{eq:replacement}), the relative weights of the observation inversions dominate those of the original estimates of the RPF.

We also use the rank histogram of the true model state (truth hereafter) as a diagnostic tool to examine the spread of the particles. Concretely, let $\hat{x}_k$ be a scalar that may be considered as an estimate of the true value $x_k^{tr}$ at time instant $k$, and $\{ \hat{x}_{k,j} \}_{j=1}^N$ an ensemble of $N$ such estimates. Then the rank $r_k$ of the truth $x_k^{tr}$ with respect to the set $\{ \hat{x}_{k,j} \}_{j=1}^N$ is obtained by sorting the magnitudes of $x_k^{tr}$ and $\hat{x}_{k,j}$ ($j = 1, \dotsb, N$) in ascending order. Collecting this information at every time step, one obtains a set of ranks $\{r_k\}_{k=0}^{S-1}$ during the assimilation time window $[0,S-1]$. A rank histogram is thus a histogram that shows the distribution of $r_k$ ($k=0,\dotsb,S-1$) during the assimilation time window. Readers are referred to, for example, \citet{Hamill2001-interpretation}, for more information of this graphical plot. In the context of particle filtering, roughly speaking, for a set of particles with reasonable variability, the corresponding rank histogram will be relatively flat, indicating that the truth is statistically indistinguishable from the particles. A U-shaped rank histogram normally indicates a spread deficiency in the particles, while a bell-shaped rank histogram indicates the opposite, i.e., over-estimated spread.

In Fig. \ref{fig:Talagrand_rank_truth_fullObv}, we show the rank histograms of the first four elements, $x_{k,i}^{tr}$ ($i=1,2,3,4$), of the truths $\mathbf{x}_k^{tr}$ ($k = 1, \dotsb, 1000$) over the whole assimilation time window in the full observation scenario. The left column shows the rank histograms of the RPF with the sample size being $20$, and the right column those of the corresponding RPF-RN (at $\beta=6$). For all of the four elements, their rank histograms in the RPF are deeply U-shaped, with the truths mostly concentrating on the edges of the histograms, meaning that the particles in the RPF substantially under-represent the variability. In contrast, the rank histograms in the RPF-RN exhibit improvements in terms of flatness (although still deeply U-shaped), meaning that better variability representations are achieved in the RPF-RN, as a by-product of residual nudging. Similar rank-histogram improvements are also observed in other observation scenarios with different sample sizes, though in some cases they may not be as significant as those shown in Fig. \ref{fig:Talagrand_rank_truth_fullObv}. 

\subsubsection{Results with different sample sizes}

Here we examine the time mean RMSEs of the RPF and the RPF-RN as functions of the sample sizes. The experiment settings are the following. We conduct the experiments in the 1/2 observation scenario, in which the observation operator is $\mathbf{H}^d$, with $d = 2$. In both filters the sample sizes $N$ are chosen from the set $\{1, 10, 20, 40, 60, 80,100, 200, 400, 600, 800,1000 \}$. In the RPF-RN, we let the noise level coefficient $\beta \in \{1, 5, 10, 15\}$.

For reference, we also investigate the performance of the EnKF with perturbed observations \citep{Burgers-analysis} under the same experiment settings. Covariance inflation \citep{Anderson-Monte} is introduced to the EnKF for all sample sizes, and covariance localization \citep{Hamill-distance} is conducted only when the sample size $N \leq 100$\footnote{Concretely, we follow the procedures in \citet{Luo2010-reply,Luo2008-spgsf1} to conduct covariance localization, in which a parameter $l_c$, called length scale, is involved in order to control the range of cut-off \citep{Hamill-distance}. On the other hand, covariance inflation is conducted by inflating a covariance matrix by a multiplicative factor $(1+\delta)^2$. In the experiment we let $\delta \in \{0:0.01:0.06\}$ and $l_c \in \{10:20:150\}$.}, following the results in Fig. 5 of \citet{bocquet2010beyond}. For conciseness, in what follows we only present the best possible results of the EnKF among the filter configurations that we have tested.

Fig. \ref{fig:varying_ensize} shows how the time mean RMSEs of, (a) the EnKF, (b) the RPF, and (c) the RPF-RN, change with the sample size. For the EnKF (Fig. \ref{subfig:EnKF_varying_ensize}), numerical results show that it diverges\footnote{Here a divergence is referred to as an event in which the RMSE of a filter at a certain time instant is larger than $10^3$.} when the sample size $N \leq 10$. For this reason, we only present the results with the sample size $N \geq 20$. As shown in Fig. \ref{subfig:EnKF_varying_ensize}, at $N=20$, the time mean RMSE of the EnKF is $4.4658$. As the sample size increases, the corresponding time mean RMSE drops rapidly until $N$ reaches $80$. After than, the time mean RMSE of the EnKF seems to enter a plateau, with the time mean RMSE being around $0.73$ and insensitive to the increase of $N$.

For the RPF (Fig. \ref{subfig:RPF_varying_ensize}), when with only 1 particle, its time mean RMSE is $5.1387$. As the sample size increases, the time mean RMSE in general tends to decrease, though there are also certain statistical fluctuations. With the sample size growing to $1000$, the time mean RMSE of the RPF reduces to $4.3195$.

\begin{figure*} 
\vspace*{2mm}
\centering

\includegraphics[width=\textwidth]{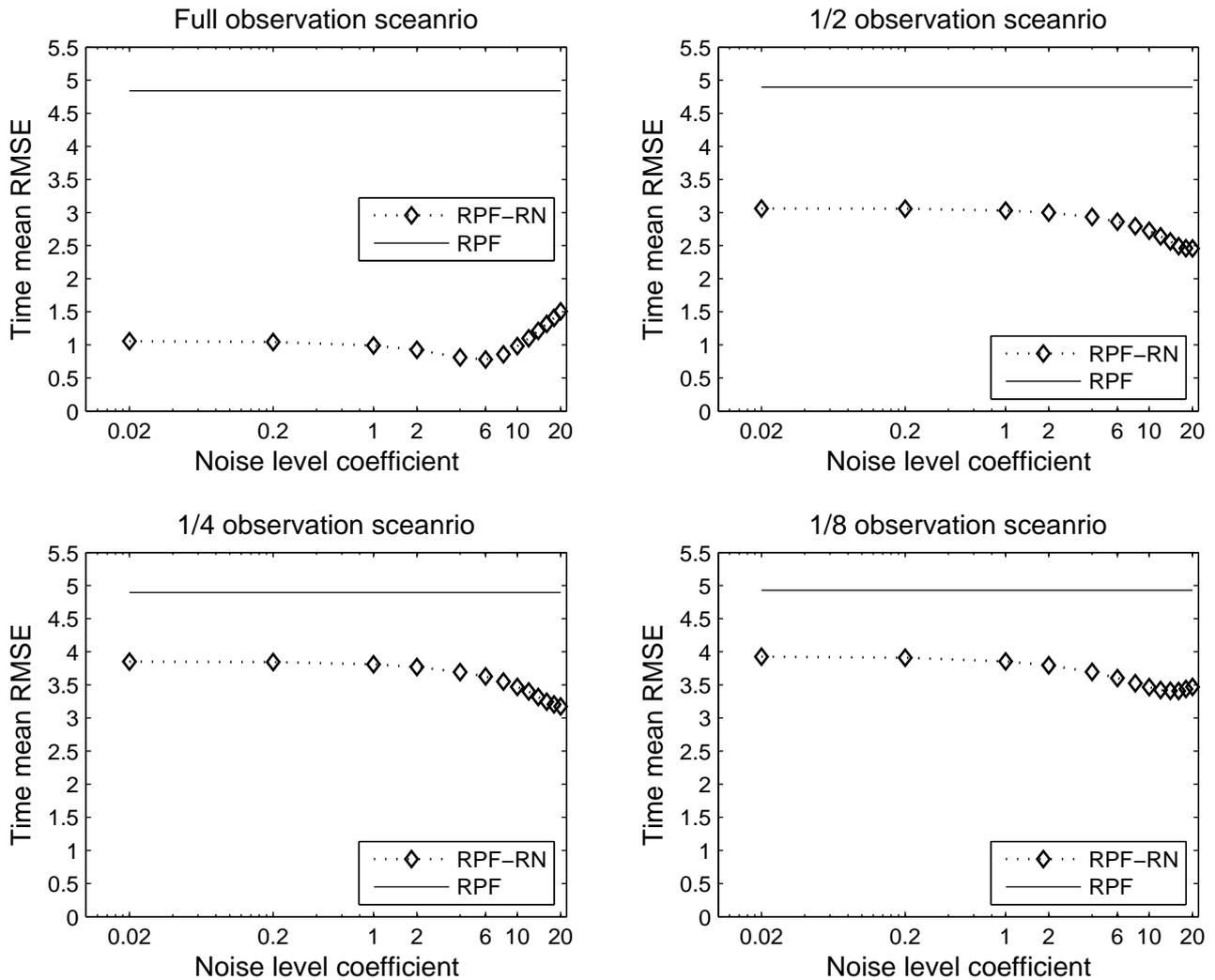}

\caption{\label{fig:DARN_RPF_rmse_vs_beta} Time mean RMSEs of the RPF-RN as functions of the noise level coefficient $\beta$ in different observation scenarios (dash-dotted lines with diamonds). For references, the corresponding time mean RMSEs of the normal RPF are also provided (solid horizontal lines).}
\end{figure*}

\begin{figure*} 
\vspace*{2mm}
\centering

\includegraphics[width=\textwidth]{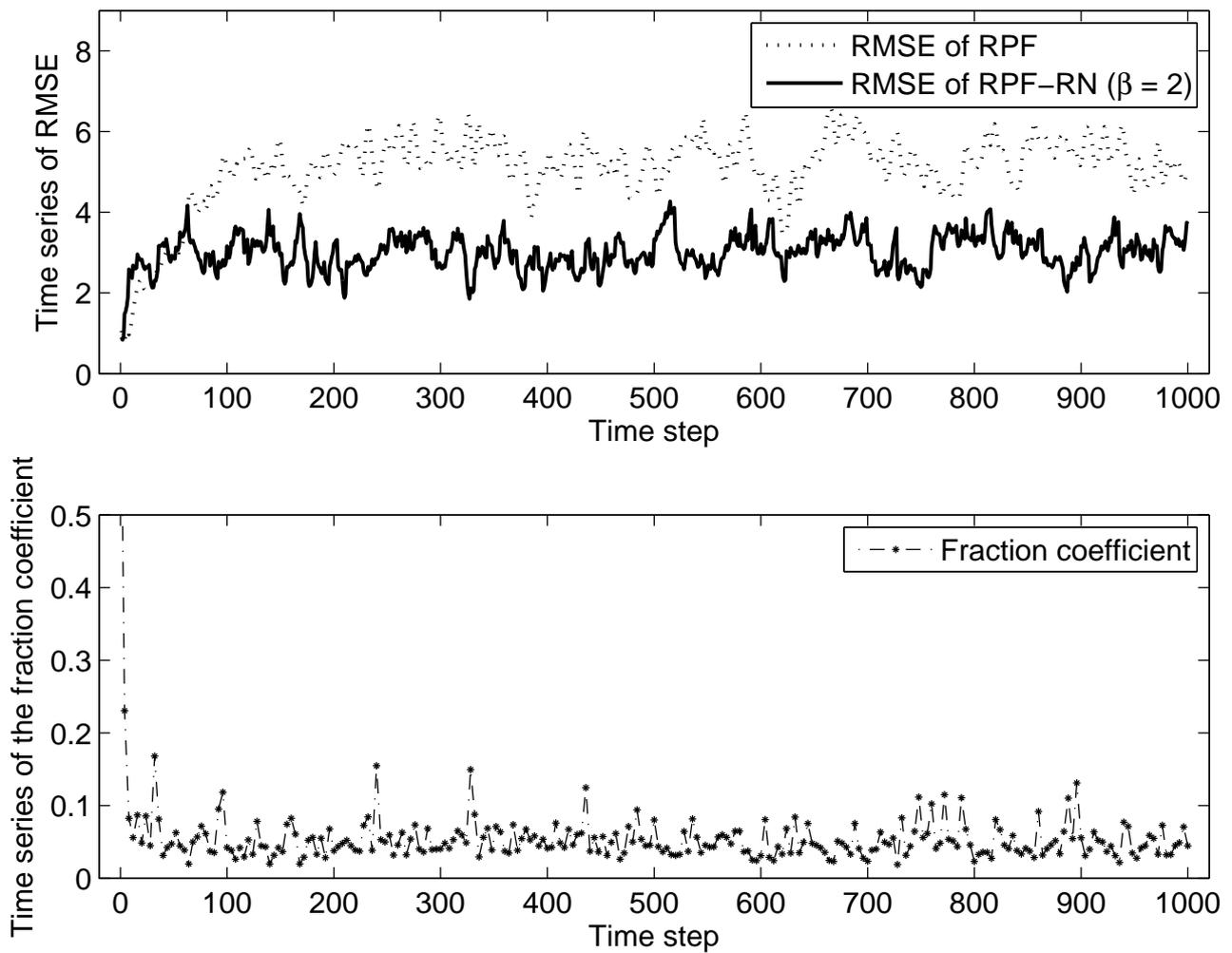}

\caption{\label{fig:RPFL96_corr_normalRPF_fraction_coefficient} Upper panel: A sample time series of the RMSEs of the RPF and RPF-RN ($\beta = 2$) in the 1/2 observation scenario; Lower Panel: Corresponding time series of the fraction coefficient of the RPF-RN ($\beta = 2$).}
\end{figure*}

\begin{figure*} 
\vspace*{2mm}
\centering

\includegraphics[width=\textwidth]{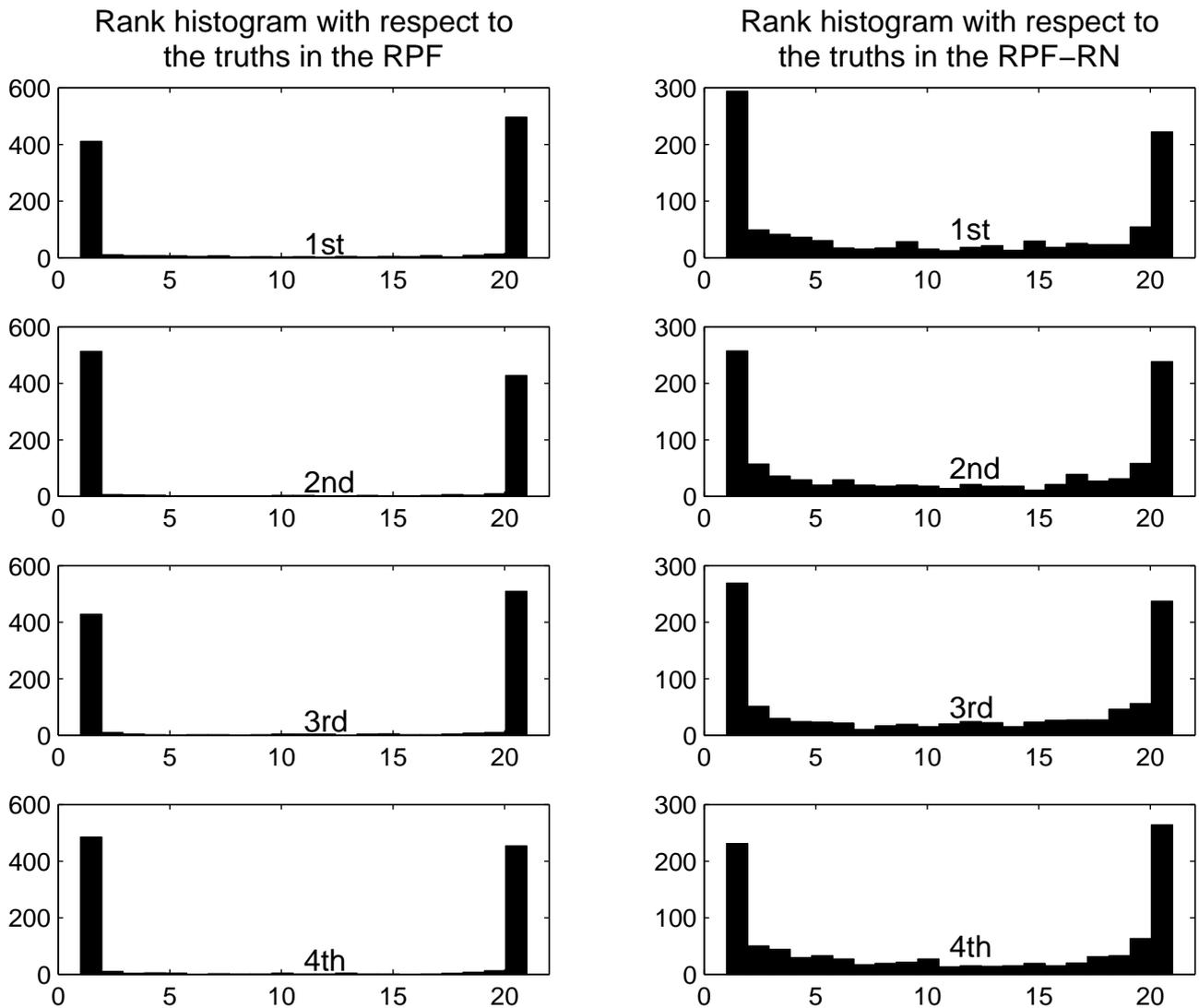}

\caption{\label{fig:Talagrand_rank_truth_fullObv} Rank histograms of the first four elements of the particles with respect to the truths in the RPF and the RPF-RN (with $\beta = 6$) in the full observation scenario.}
\end{figure*}

\begin{figure*} 
\vspace*{2mm}
\centering

\subfigure[Results of the EnKF]{ \label{subfig:EnKF_varying_ensize}
\includegraphics[scale = 0.5]{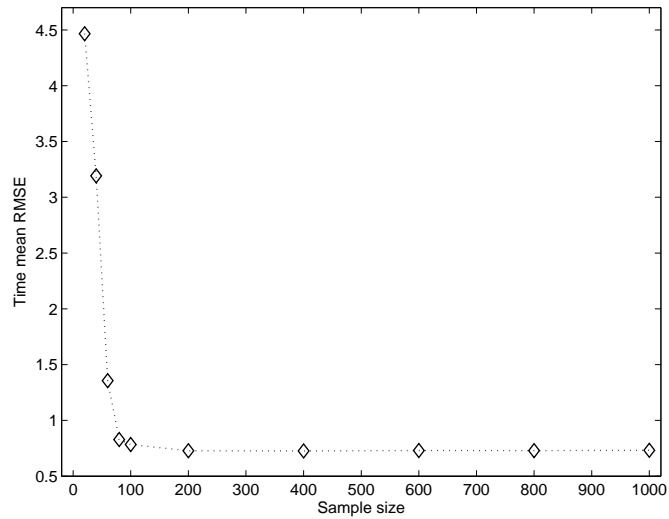}
}

\subfigure[Results of the RPF]{ \label{subfig:RPF_varying_ensize}
\includegraphics[scale = 0.5]{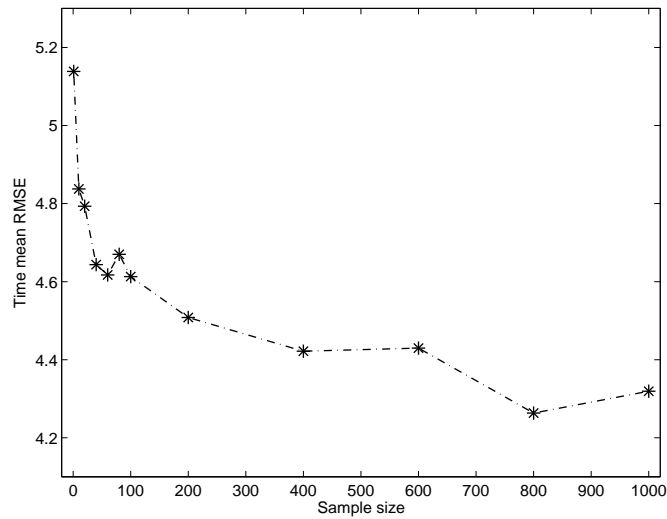}
}

\subfigure[Results of the RPF-RN]{ \label{subfig:RPF_RN_varying_ensize}
\includegraphics[scale = 0.5]{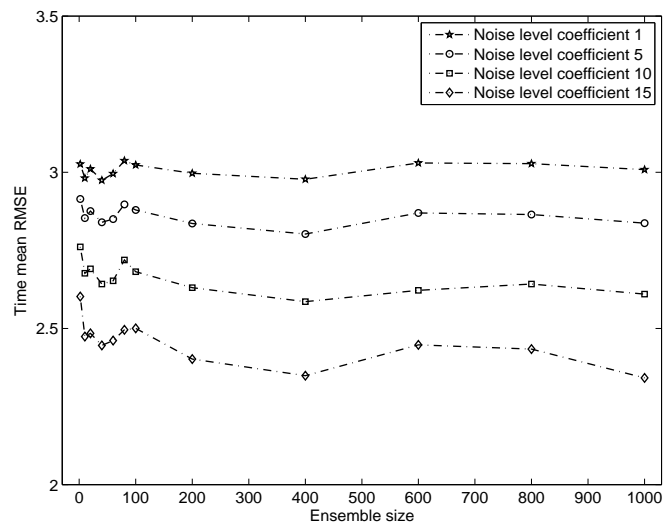}
}

\caption{\label{fig:varying_ensize} Time mean RMSEs of (a) the EnKF; (b) the RPF; and (c) the RPF-RN, as functions of the sample size in the 1/2 observation scenario. Note that in the EnKF, filter divergence is spotted with sample size $N=1$ and $N=10$ so that the results of the EnKF are reported from $N=20$.}
\end{figure*}

Fig. \ref{subfig:RPF_RN_varying_ensize} shows the corresponding time mean RMSEs of the RPF-RN with different $\beta$. The following phenomena are observed. (1) For a fixed sample size, the time mean RMSE decreases as $\beta$ increases from $1$ to $15$; (2) The RPF-RN with different $\beta$ exhibits similar response to the change of the sample size. When the sample size is lower than $100$, the time mean RMSEs of the RPF-RN follow a U-turn behaviour, achieving the minimum values somewhere between sample size $1$ and $100$. For sample sizes larger than $100$, the time mean RMSEs of the RPF-RN also seem to follow a U-turn behaviour, achieving their minimum values around sample size $600$. A possible explanation of these phenomena is that changing the sample size has an effect on the residual norm $\Vert \hat{\mathbf{r}}^a_k \Vert_{\mathbf{R}_k}$. This in effect is equivalent to changing the $\beta$ value in Eq. (\ref{eq:fraction_coefficient}) with a fixed residual norm $\Vert \hat{\mathbf{r}}^a_k \Vert_{\mathbf{R}_k}$, and may thus cause the U-turn behaviour, as have already been observed in Fig. \ref{fig:DARN_RPF_rmse_vs_beta}.

Comparing Figs. \ref{subfig:RPF_varying_ensize} and \ref{subfig:RPF_RN_varying_ensize}, it is clear that, with the above specific experiment settings, the RPF-RN with $\beta \in \{1, 5, 10, 15\}$ systematically outperforms the RPF in terms of time mean RMSE. Even with the sample size of $1000$, the estimate of the RPF is still less informative than the observation inversion (cf. the time mean RMSE at $\beta =0.02$ in the upper right panel of Fig. \ref{fig:DARN_RPF_rmse_vs_beta}). As a result, in Fig. \ref{subfig:RPF_RN_varying_ensize} one can see that the estimate of the RPF-RN with only 1 particle is still (much) better than that of the RPF with 1000 particles. This shows that it is possible, in certain circumstances, for the RPF-RN with a relatively small sample size to achieve better filter performance than that of the normal RPF with a substantially larger sample size, similar to the result reported in \citet{vanLeeuwen2010-nonlinear}. This conclusion, however, should only be interpreted in conjunction with the above experiment settings.

Finally, a comparison between the RPF-RN and the EnKF shows that, when the sample size is relatively small, say, $N \leq 40$, the RPF-RN tends to outperform the EnKF. With a larger sample size, though, the EnKF may perform much better than the RPF-RN instead. We stress that the conclusion that the RPF-RN performs better than the EnKF (with perturbed observation) for a relatively small ensemble size may depend on the experiment setting. For instance, if one replaces the EnKF by the ensemble adjustment Kalman filter (EAKF) \citep{Anderson-ensemble}, then with the sample size $20$ the EAKF may outperform the RPF-RN instead (see, for example, the numerical results in \citealt{Luo2012-residual}). A related question is then when it is recommended to use the particle filter with residual nudging (PF-RN), instead of the EnKF. In our opinion, advantages in using the PF-RN may include that its performance appears more robust with relatively small sample sizes (in this aspect one may wish to compare the numerical results in different scenarios that are presented in this study and those in \citealt{Luo2012-residual}), and that there is no need to tune the intrinsic filter parameters in the EnKF, i.e., the covariance inflation factor and the length scale of covariance localization. As shown in \citet{Luo2012-residual}, in certain circumstances the EnKF may diverge for some combinations of the covariance inflation factor and the length scale of covariance localization. Therefore, in practice if one is only able to afford a small sample size, it might be worth to run a PF-RN first, and then, if possible (and desirable), use the PF-RN estimate as the baseline to see if it would be better to use the EnKF instead. On the other hand, we envision that there is still space of improvement for the PF-RN in the future. One possibility is to equip the PF-RN with a better importance density to further mitigate the effect of weight degeneracy, which may be done by, for instance, combining the equal-weight particle filter \citep{vanLeeuwen2010-nonlinear,ades2012exploration} with residual nudging.

\subsubsection{Results with different assimilation frequencies and observation noise covariances } \label{subsubsec:frquency}
Here we examine the effects of the assimilation frequency and the observation noise covariance matrix on the performance of the RPF and RPF-RN. To this end, we vary the assimilation frequency, and choose to assimilate the observations for every $S_a$ step(s), with $S_a \in \{1, 2, 4, 6, 8, 10, 12\}$. For convenience, we call $S_a$ the assimilation step when it causes no confusion. To examine the effect of the observation noise covariance matrix, we assume that the covariance matrix $\mathbf{R}_k$ is of the form $\gamma \mathbf{I}$, where $\mathbf{I}$ is the identity matrix with a suitable dimension, and $\gamma$ a positive scalar. As a result, the variances are $\gamma$ for all measurements in an observation vector, while the cross-variances are all zero. In the experiment we choose the variance $\gamma$ from the set $\{0.01,0.1,1,10\}$. The relatively large value of $\gamma$ at $10$ is used to represent the scenario in which the quality of the observations is relatively poor. Here we assume that we know $\gamma$ precisely, while for the experiment in the next sub-section (Section \ref{subsubsec:misspecified_obs}), we will consider the case in which $\gamma$ is mis-specified. In the experiment we consider both the 1/2 and 1/40 observation scenarios. In the latter case only the first element of the model state is observed (equivalent to setting $d = 40$ in Eq. (\ref{eq:L96_obs})), a scenario in which the filters may be subject to divergences. Other experiment settings are as following. The sample size $N$ is $20$ for both the RPF and the RPF-RN (unless otherwise mentioned). In the RPF-RN we set $\beta = 0.02$, which is a relatively small value chosen to enhance the stability of the RPF-RN (see the discussion in Section \ref{sec:RN discussion}).

\begin{figure*} 
\vspace*{2mm}
\centering

\includegraphics[width=\textwidth]{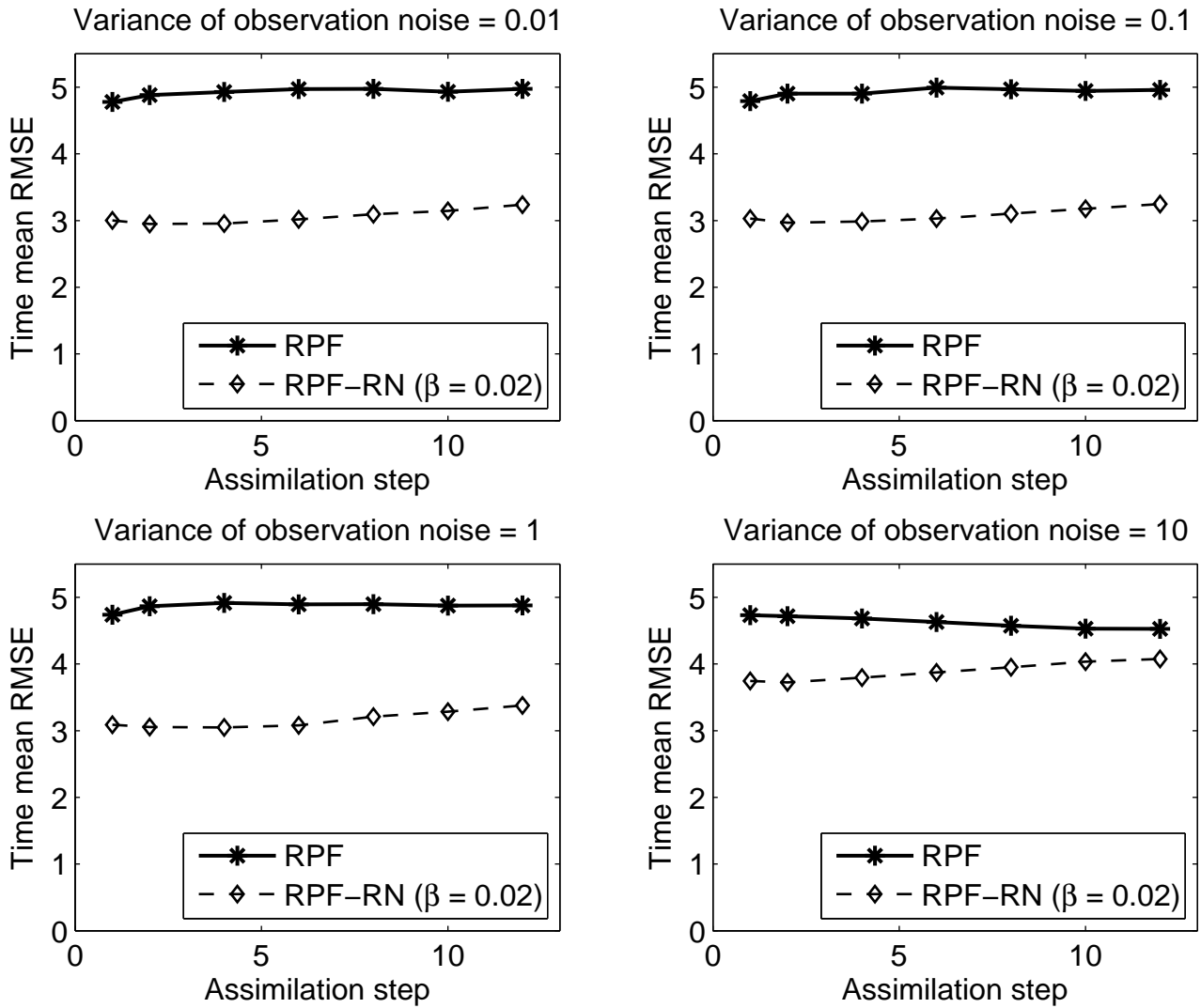}

\caption{\label{fig:RPF_L96_varying_asmStep_vs_obvLvl_obvSkip2} Time mean RMSEs of the RPF and RPF-RN with different assimilation steps and observation noise variances in the 1/2 observation scenario.}
\end{figure*}

Fig. \ref{fig:RPF_L96_varying_asmStep_vs_obvLvl_obvSkip2} reports the performance of the RPF with different $S_a$ and $\gamma$ in the 1/2 observation scenario (solid lines with asterisks). When $\gamma$ is relatively small, say $\gamma = 0.01, 0.1$ and $1$ (upper left, upper right, and lower left panels, respectively), the time mean RMSE of the RPF is the smallest at $S_a = 1$, and tends to increase as $S_a$ grows. For a sufficiently large $S_a$ (say at $S_a = 6$), though, further increasing $S_a$ does not significantly change the performance of the filter. Interestingly, at $\gamma = 10$ (lower right panel), the RPF with a larger $S_a$ tends to have better performance than the RPF with a smaller $S_a$. In addition, when $S_a$ is relatively large, say $S_a = 12$, the RPF at $\gamma = 10$ performs better than the RPF at other smaller $\gamma$ values. For comparison, we also show the corresponding performance of the RPF-RN ($\beta = 0.02$) in the same figure (dash lines with diamonds). For a fixed $\gamma$, the time mean RMSE of the RPF-RN is slightly U-shaped as $S_a$ changes, and it tends to achieve its minimum with $S_a > 1$. On the other hand, for a fixed $S_a$, the time mean RMSE of the RPF-RN tends to increase as $\gamma$ increases. In all the tested cases, the time mean RMSEs of the RPF-RN are lower than the corresponding values of the normal RPF.

It might not be consistent with our intuition to see that a PF with a larger assimilation step $S_a$ and worse observation quality (in the sense of having a larger $\gamma$) has better performance. In our opinion, though, this might be explained from the point of view of the effects of $S_a$ and $\gamma$ on the effective sample size (ESS). Let $\{ \mathbf{x}_{k,i} \}_{i=1}^N$ be the set of particles that are associated with the weights $\{ w_{k,i} \}_{i=1}^N$ after applying the weight update formula Eq. (\ref{eq:weight_update_PF}), but before conducting re-sampling (if any). The ESS, defined as $1/ (\sum\limits_{i=1}^N w_{k,i}^2)$ \citep{liu1995blind}, can be used as an indicator of the degree of weight collapse at time instant $k$ in a particle filter. One can also define time mean ESS in a way similar to that in defining the time mean RMSE (see the text below Eq. (\ref{eq:def_of_rmse})).

In terms of time mean ESS and information contents of incoming observations, large $S_a$ and $\gamma$ have both positive and negative effects on the filter performance. A relatively large assimilation step $S_a$ means that there are more model integration steps in between two successive observations. For the relatively small sample size $N=20$, re-sampling is often performed, after which the re-sampled particles have uniform weights. These uniform weights are then carried to the subsequent model integration steps, until they are updated with the next incoming observation. As a result, a larger assimilation step $S_a$ implies that, on the one hand, the time mean ESS of the particle filter tends to be larger, while on the other, less information contents of the observations are assimilated. Similarly, a larger $\gamma$ tends to make the weights of the particles more uniform, which in turn increases the time mean ESS. On the other hand, though, observations with a larger $\gamma$ contain more uncertainties and less information about the underlying model state. Therefore in our opinion, the reported behaviour of both filters in Fig. \ref{fig:RPF_L96_varying_asmStep_vs_obvLvl_obvSkip2} may largely result from the combined positive and negative effects in choosing $S_a$ and $\gamma$. For verification, in Table \ref{table:ess_of_normal_RPF} we show the time mean ESS of both the RPF and RPF-RN ($\beta = 0.02$) with different combinations of $S_a$ and $\gamma$ in the 1/2 observation scenario. As can be seen there, the time mean ESS of both filters indeed tend to increase as $S_a$ and/or $\gamma$ increase(s).

\begin{figure*} 
\vspace*{2mm}
\centering

\includegraphics[width=\textwidth]{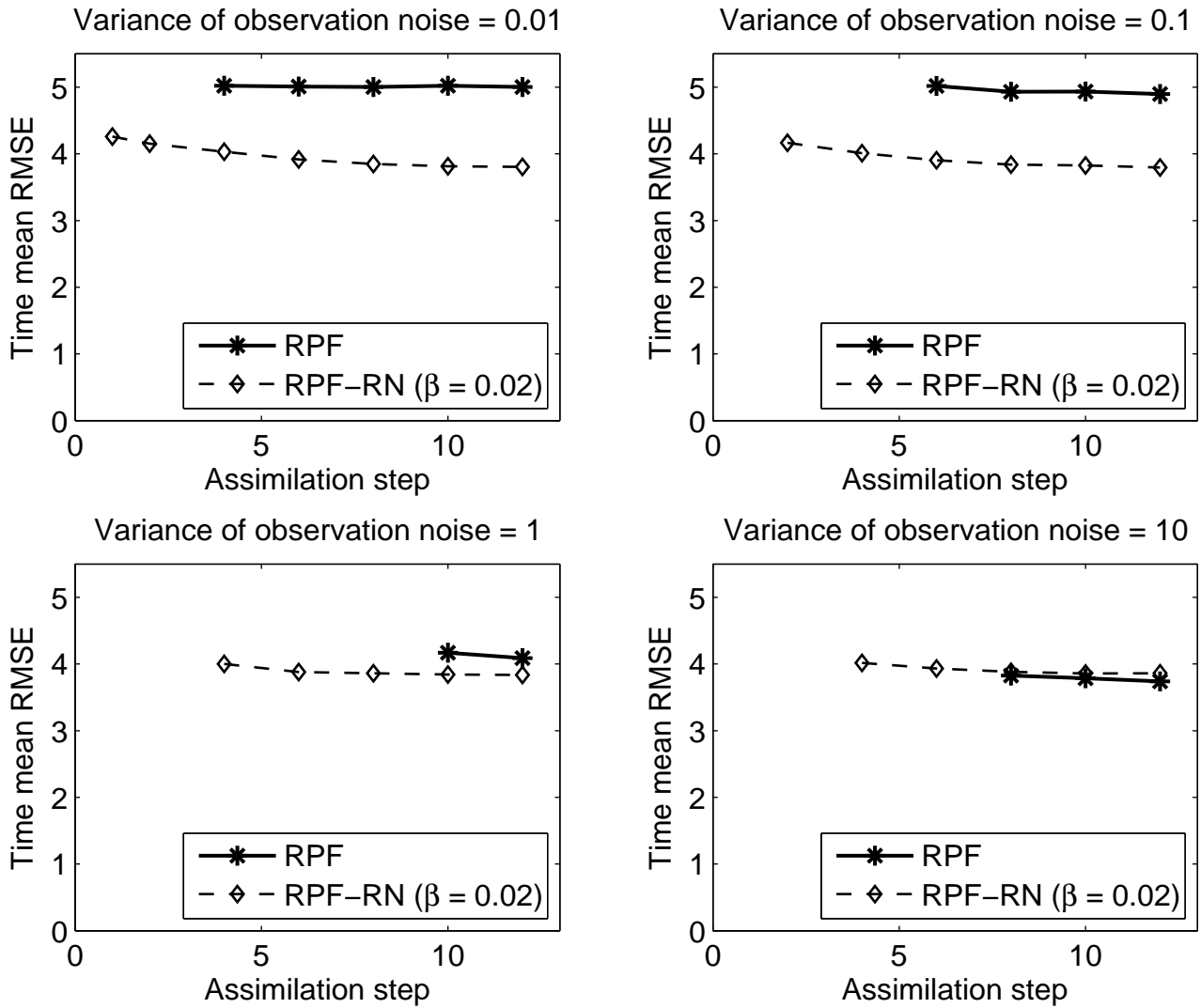}

\caption{\label{fig:RPF_L96_varying_asmStep_vs_obvLvl_obvSkip40} As in Fig. \ref{fig:RPF_L96_varying_asmStep_vs_obvLvl_obvSkip2}, but it is now in the 1/40 observation scenario.}
\end{figure*}

Fig. \ref{fig:RPF_L96_varying_asmStep_vs_obvLvl_obvSkip40} shows the performance of the RPF (solid lines with asterisks) and RPF-RN (dash lines with diamonds) in the 1/40 observation scenario. Both filters exhibit behaviour similar to that in Fig. \ref{fig:RPF_L96_varying_asmStep_vs_obvLvl_obvSkip2}, e.g., the filters may have better performance with a larger $S_a$ and/or $\gamma$ (when there is no filter divergence). 
An examination of the time mean ESS of both filters shows that they are close to the time mean ESS reported in Table \ref{table:ess_of_normal_RPF}, and are thus not shown for conciseness.

Compared with Fig. \ref{fig:RPF_L96_varying_asmStep_vs_obvLvl_obvSkip2}, there are also a few differences in Fig. \ref{fig:RPF_L96_varying_asmStep_vs_obvLvl_obvSkip40}. One is that, unlike the situation in the 1/2 observation scenario, filter divergences are spotted in both the RPF and RPF-RN in certain circumstances. Accordingly, when a filter divergence is spotted, there will be no RMSE value plotted in the corresponding place in Fig. \ref{fig:RPF_L96_varying_asmStep_vs_obvLvl_obvSkip40}. Following this setting, for instance, the upper right panel (with $\gamma = 0.1$) of Fig. \ref{fig:RPF_L96_varying_asmStep_vs_obvLvl_obvSkip40} shows that the normal RPF diverges at $S_a = 1$, $2$ and $4$, while the RPF-RN diverges at $S_a = 1$ only. In terms of stability against divergence, the results in Fig. \ref{fig:RPF_L96_varying_asmStep_vs_obvLvl_obvSkip40} show that the RPF-RN ($\beta = 0.02$) tends to be more stable than the normal RPF at different $\gamma$ values. In addition, when $\gamma$ is relatively small, say, $\gamma = 0.01$, $0.1$ and $1$, the RPF-RN still tends to perform better than the normal RPF in terms of filter accuracy. However, at $\gamma = 10$ and with $S_a = 8$, $10$ and $12$, the time mean RMSEs of the RPF-RN become larger than those of the RPF. This is largely because in such cases the RPF achieves reasonable performance, while with a relatively small value $\beta = 0.02$, the RPF-RN tends to rely excessively on the observation inversion (cf. Eqs. (\ref{eq:fraction_coefficient}) and (\ref{eq:replacement})). The RPF-RN can also have time mean RMSEs that are very close to those of the normal RPF by increasing $\beta$ to, for instance, $10$. This choice, however, may make the RPF-RN less stable at smaller $S_a$ values. This serves as an example to show that $\beta$ has an impact on the trade-off between a filter's potential accuracy and stability.

\begin{figure*} 
\vspace*{2mm}
\centering

\includegraphics[width=\textwidth]{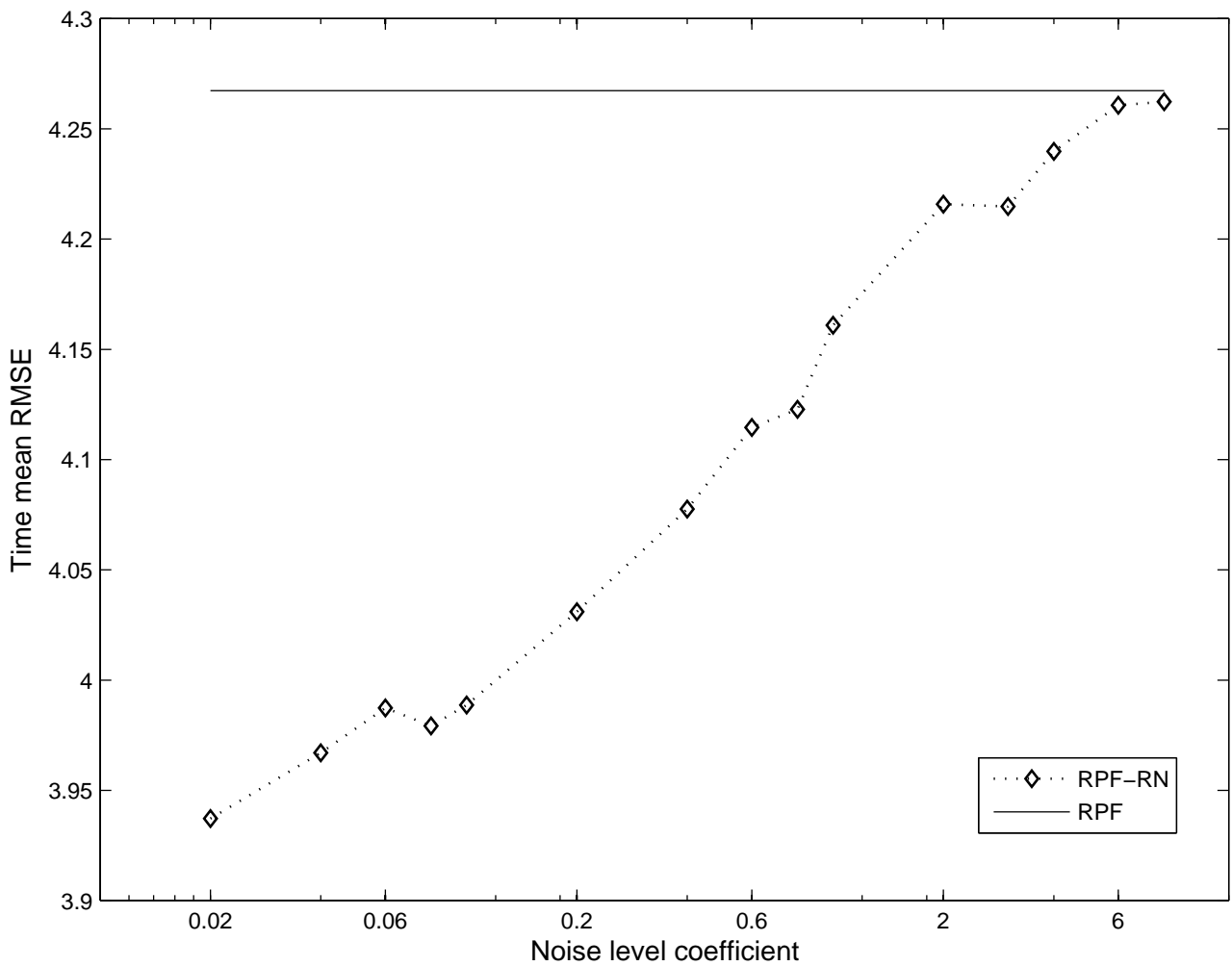}

\caption{\label{fig:RPF_L96_varying_beta_obvSkip40_Sa12_obvLv10_ensize5} Time mean RMSEs of the RPF and RPF-RN in the 1/40 observation scenario. The experiment settings are: the sample size $N = 5$, the assimilation step $S_a = 12$, and the observation noise variance $\gamma = 10$. In the RPF-RN $\beta \in \{0.02:0.02:0.1, 0.2:0.2:1, 2, 3, 4, 6, 8\}$.}
\end{figure*}

An additional remark regarding the relatively superior performance of the normal RPF in the lower right panel of Fig. \ref{fig:RPF_L96_varying_asmStep_vs_obvLvl_obvSkip40} is the following. The effective dimension of the L95 model, in terms of the {\color{black}{``Kaplan-York'' dimension}} \citep{ruelle1989chaotic}, is about $27.1$ \citep{Lorenz-optimal}, while the time mean ESS of the normal RPF at $S_a = 8$, $10$ and $12$ are around $18$, not quite far away from the fractal dimension. In such cases, the subspace spanned by the particles of the normal RPF may capture the state space features of the L95 model reasonably well. As a result, in this specific context, the information contents of the observations may not be very influential on the estimation accuracy of the normal RPF (but may still be useful in terms of filter stability against divergence). If the subspace spanned by the particles is a less proper representation of the state space, we expect that the information contents of the observations may become more important to the filter performance. To this end, we conduct one more experiment in the 1/40 observation scenario, in which we let $S_a = 12$ and $\gamma = 10$, but reduce the sample size $N$ of the normal RPF to $N=5$. For comparison, we also examine the performance of the RPF-RN with different noise level coefficients $\beta \in \{0.02:0.02:0.1, 0.2:0.2:1, 2, 3, 4, 6, 8\}$. The time mean RMSEs of the RPF (solid horizontal line) and RPF-RN (dash-dotted line with diamonds), as functions of $\beta$, are shown in Fig. \ref{fig:RPF_L96_varying_beta_obvSkip40_Sa12_obvLv10_ensize5}. Compared with the lower right panel of Fig. \ref{fig:RPF_L96_varying_asmStep_vs_obvLvl_obvSkip40}, the performance of both filters deteriorates. However, in all the tested cases the RPF-RN performs better than the normal RPF. In addition, the RPF-RN tends to have a lower time mean RMSE with a smaller $\beta$, meaning that the RPF-RN has better performance when it relies more on the observations in residual nudging.

\subsubsection{Results with inaccurately specified models and observation systems} \label{subsubsec:misspecified_obs}
Finally we examine the performance of the RPF and the RPF-RN in the presence of errors in specifying the dynamical model and the observation system. For convenience of discussion, we confine ourselves to the 1/2 and 1/40 observation scenarios. In the 1/2 observation scenario, we assume that in the experiments the forcing term $F$ in Eq.~(\ref{eq:LE98}) and the observation error covariance $\mathbf{R}_k$ are possibly mis-specified. The true value of $F$ is 8, and the true observation error covariance $\mathbf{R}_k$ is $\mathbf{I}_{20}$. In the experiments we let the value of $F$ be chosen from the set $\{ 4, 6, 8, 10, 12 \}$, and $\mathbf{R}_k$ be in the form of $\gamma \mathbf{I}_{20}$, with the observation noise variance $\gamma \in \{ 0.25, 0.5, 1, 2, 5, 10\}$. Note that in the RPF-RN, $\mathbf{R}_k$ is not only used to update the weights of the particles as in Eq.~(\ref{eq:weight_update_PF}), but also used to compute the fraction coefficient $c_k$ in residual nudging (cf. Eq.~(\ref{eq:fraction_coefficient})). We let the sample size $N = 20$ and the assimilation step $S_a = 4$ in both filters, and the noise level coefficient $\beta = 1$ in the RPF-RN.

\begin{figure*} 
\vspace*{2mm}
\centering

\includegraphics[width=\textwidth]{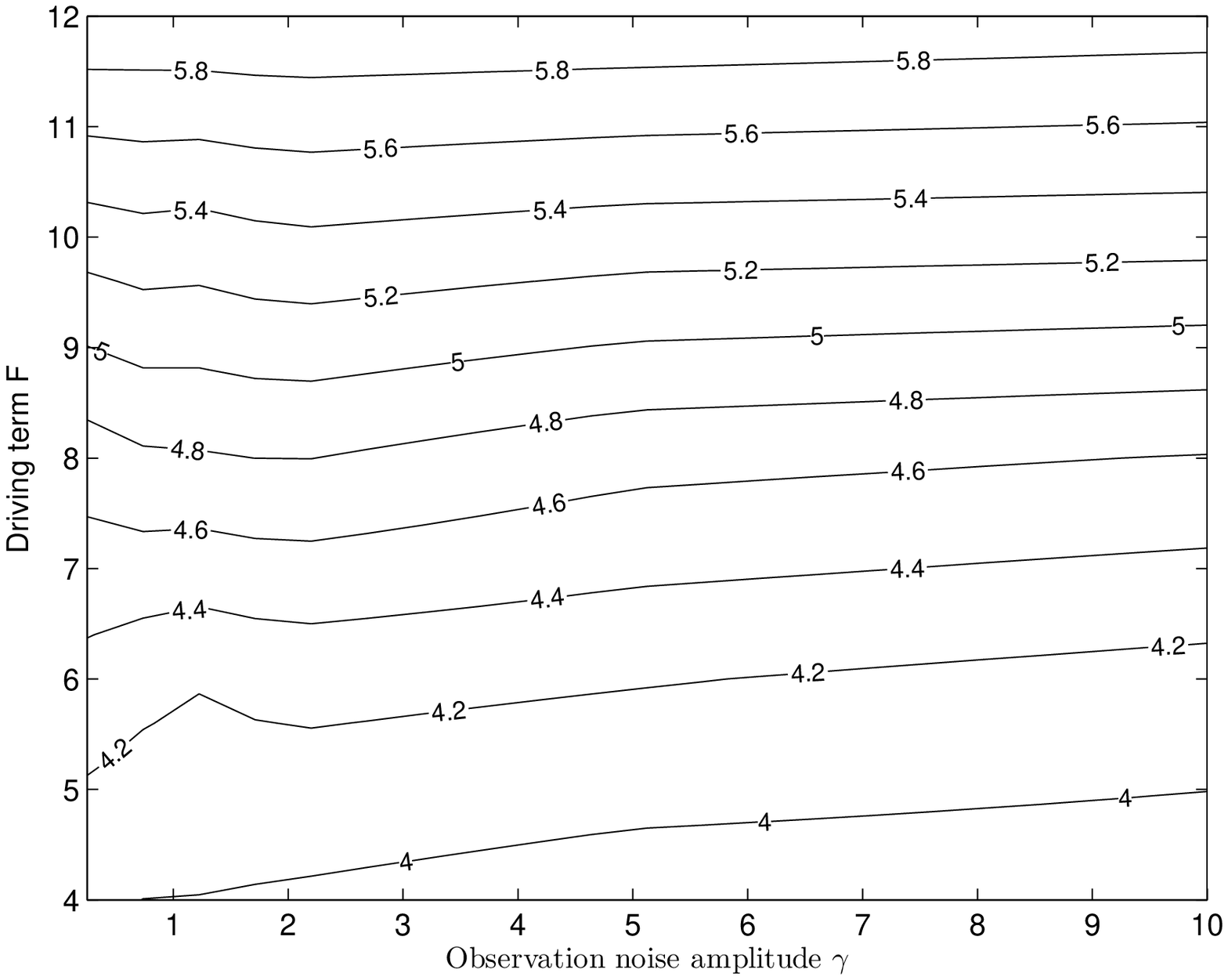}

\caption{\label{fig:Normal_RPF_F_vs_gamma} Time mean RMSE of the RPF as a function of the driving term $F$ and the observation noise variance $\gamma$ in the 1/2 observation scenario.}
\end{figure*}

Fig. \ref{fig:Normal_RPF_F_vs_gamma} shows the contour plot of the time mean RMSE of the RPF, with respect to the values of the forcing term $F$ and the observation noise variance $\gamma$, in the 1/2 observation scenario. For a fixed $\gamma$, the time mean RMSE of the RPF tends to increase as $F$ grows. On the other hand, for a fixed $F$, the time mean RMSE tends to decrease as $\gamma$ grows (which may also be explained based on the arguments in Section \ref{subsubsec:frquency}), with the decrement rates becoming smaller at larger $F$ values.

\begin{figure*} 
\vspace*{2mm}
\centering

\includegraphics[width=\textwidth]{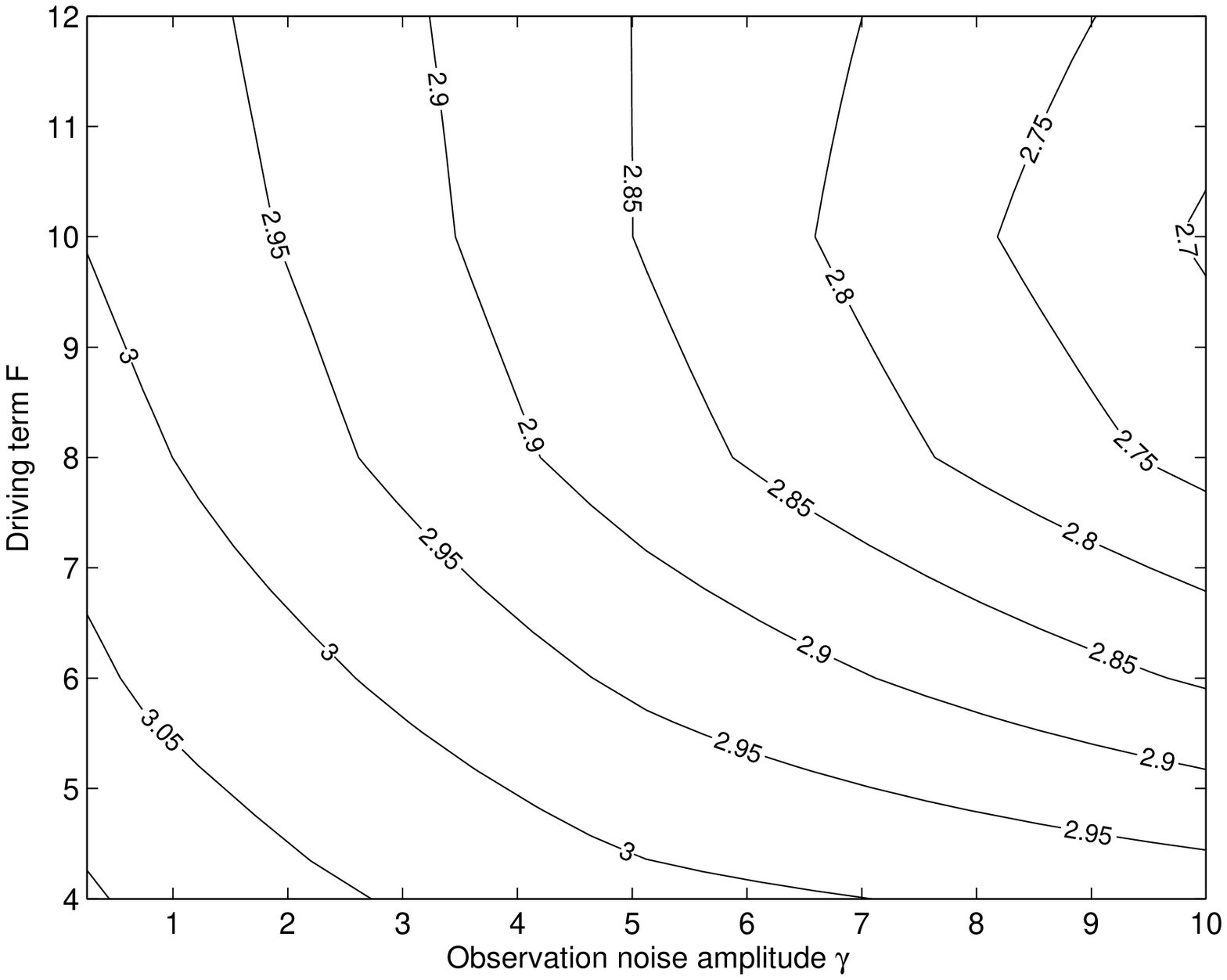}

\caption{\label{fig:DARN_RPF_F_vs_gamma} Time mean RMSE of the RPF-RN (with $\beta = 1$) as a function of the driving term $F$ and the observation noise variance $\gamma$ in the 1/2 observation scenario.}
\end{figure*}

For comparison, Fig. \ref{fig:DARN_RPF_F_vs_gamma} depicts the corresponding contour plot of the time mean RMSE of the RPF-RN in the 1/2 observation scenario. There appears to be a ``sink'' around the point $(F=10,\gamma = 10)$. Along a fixed direction, the further away from the sink, the larger the time mean RMSE tends to be. Comparing Figs. \ref{fig:Normal_RPF_F_vs_gamma} and \ref{fig:DARN_RPF_F_vs_gamma}, one can see that the RPF-RN again outperforms the RPF in all tested cases. In fact, even the largest time mean RMSE of the RPF-RN (around the lower left corner of Fig. \ref{fig:DARN_RPF_F_vs_gamma}) is still lower than the best time mean RMSE of the RPF (around the lower right corner of Fig. \ref{fig:Normal_RPF_F_vs_gamma}).

The experiment settings of the 1/40 observation scenarios are almost the same as those in the 1/2 observation scenario, except that the assimilation step $S_a$ of both filters becomes $12$. With fewer measurements in an observation vector, filter divergences are also spotted in some cases. Therefore, instead of presenting the contour plots, we choose to directly report the time mean RMSEs of both filters in Table \ref{table:time_mean_RMSE}, in which filter divergences are marked by ``Div'' in relevant places. The results there show that, for a fixed $F$, the time mean RMSEs of both filters tend to decrease as $\gamma$ grows. On the other hand, for a fixed $\gamma$, the time mean RMSE of the RPF tends to grow with $F$ when $\gamma$ is relatively small (say, at $\gamma = 0.25$), and exhibits slightly U-shaped behaviour when $\gamma$ is relatively large (say, at $\gamma = 10$). The time mean RMSE of the RPF-RN also exhibits similar behaviour. Overall, the RPF-RN ($\beta = 1$) tends to perform better than the RPF, although compared to the results in Figs. \ref{fig:Normal_RPF_F_vs_gamma} and \ref{fig:DARN_RPF_F_vs_gamma}, the gap between the RPF and RPF-RN ($\beta = 1$) is clearly narrowed. Both filters diverge in all cases with $F=12$, and two other cases at $(F=10,\gamma = 0.5)$ and $(F=10,\gamma = 1)$. However, numerical experiments (results not reported) show that in this case one can improve the stability of the RPF-RN by reducing $\beta$ to some smaller value, say, $0.02$. 

\section{Conclusion} \label{sec:conclusion}

In this work we considered an observation-space based auxiliary technique, called residual nudging, to enhance the performance of the particle filter. The main idea of residual nudging is to monitor, and if necessary, adjust the residual norm of a state estimate so that it does not exceed a pre-specified threshold. We suggested a rule to choose the threshold, and proposed a method to do the possible adjustment in case of linear observations. For demonstration, we used the regularized particle filter (RPF) to conduct data assimilation in an AR1 model and the 40-dimensional Lorenz 95 model. The experiment results showed that the RPF with residual nudging (RPF-RN) outperforms the normal RPF in terms of filter accuracy and/or stability again divergence, especially when the normal RPF performs poorly.

A problem that is not fully addressed in this work is the nonlinearity in the observation operator. We envision that residual nudging would be still applicable, with the same rationale in choosing the pre-specified threshold $\beta \sqrt{p}$ as discussed in the text below Eq.~(\ref{eq:residual_norm}). When the observation operator $\mathcal{H}_k$ is continuous with respect to the model state, and there exists an observation inversion $\mathbf{x}^o_k$ such that $\mathcal{H}_k(\mathbf{x}^o_k) = \mathbf{y}^o_k$, 
then the objective of residual nudging can be achieved, i.e., there exists a fraction coefficient $c_k \in [0,1]$ such that the residual norm of the estimate $\breve{\mathbf{x}}_k^a$ obtained through Eq. (\ref{eq:replacement}) is no larger than $\beta \sqrt{p}$. With nonlinearity, though, it may become more complicated in finding the estimate $\breve{\mathbf{x}}_k^a$. While possible strategies in handling nonlinearity were mentioned in Section \ref{subsec:RN}, how to implement them in numerically efficient ways will be investigated in the future.

\ack We thank Dr. M. Bocquet, Dr. C. Snyder and two anonymous reviewers for their constructive and inspiring comments and suggestions. We have also benefited from the useful discussions with Dr. Geir N{\ae}vdal and Dr. Andreas Stordal at IRIS. Luo acknowledges partial financial support from the Research Council of Norway and industrial partners through the project "Transient well flow modelling and modern estimation techniques for accurate production allocation".

\appendix
\numberwithin{equation}{section}
\section{Outline of the regularized particle filter} \label{sec:algorithm}
Instead of using importance re-sampling as in the conventional bootstrap particle filter \citep{Gordon1993}, the RPF employs an alternative way to tackle the problem of particle degeneracy based on kernel density estimation (KDE, see, for example, \citealp{Silverman1986}). The idea is to construct a continuous pdf based on the original particles and their associated weights. This continuous pdf is treated as an approximation of the underlying pdf of the true model state, and is used to draw a new set of particles that is different from the original one almost surely \citep[ch. 12]{Doucet2001-sequential}.

For illustration, suppose that at the $k$-th assimilation cycle there is a set of $N$ (original) particles $\{ \mathbf{x}_{k,i} \}_{i=1}^N$, together with the corresponding weights $\{ w_{k,i} \}_{i=1}^N$. As a result,
\begin{linenomath*}
\begin{equation}
\mathbf{S}_k = \dfrac{1}{N-1} \left[ \sqrt{w_{k,1}} ( \mathbf{x}_{k,1} - \hat{\mathbf{x}}_k  ),\dotsb, \sqrt{w_{k,N}} ( \mathbf{x}_{k,N} - \hat{\mathbf{x}}_k  ) \right] \, ,
\end{equation}
\end{linenomath*}
is a square root of the weighted sample covariance with respect to the particles $\{ \mathbf{x}_{k,i} \}_{i=1}^N$, where
\begin{linenomath*}
\begin{equation}
\hat{\mathbf{x}}_k = \sum\limits_{i=1}^N w_{k,i} \, \mathbf{x}_{k,i}
\end{equation}
\end{linenomath*}
is the weighted sample mean.

The continuous pdf to be constructed is then expressed in the form of \citep[ch. 12]{Doucet2001-sequential}
\begin{linenomath*}
\begin{equation}
\tilde{p} (\mathbf{x}_k) = \sum\limits_{i=1}^N  w_{k,i} K \left( \dfrac{\mathbf{x}_k -\mathbf{x}_{k,i}}{h} \right) \, ,
\end{equation}
\end{linenomath*}
where $K(\bullet)$ is the kernel function and $h$ is a scalar parameter called the bandwidth \citep{Silverman1986}. For the RPF implemented in this study, we use the Gaussian kernel and choose the bandwidth $h$ according to the following rule (cf. \citealp[Eq. (12.2.7)]{Doucet2001-sequential}):
\begin{linenomath*}
\begin{subequations}
\begin{align}
& A = (\dfrac{4}{n + 2})^{1/(n + 4)} \, , \\
& h = A \, N^{-1/(n + 4)} \, ,
\end{align}
\end{subequations}
\end{linenomath*}
where $n$ is the dimension of $\mathbf{x}_k$.

    \newenvironment{myindentpar}[1]%
     {\begin{list}{}%
             {\setlength{\leftmargin}{#1}}%
             \item[]%
     }
     {\end{list}}

The main procedures of the RPF implemented in this study are summarized below, largely following the style in \citet[Algorithm 6]{Arulampalam2002}.
\begin{itemize}
\item Prediction step: FOR i = 1 to N

	Draw a prior sample $\mathbf{x}_{k,i}^b$ from the transition pdf $p \left( \mathbf{x}_{k} | \tilde{\mathbf{x}}^a_{k-1,i} \right)$, and assign the weight $\tilde{w}_{k-1,i}$ of $\tilde{\mathbf{x}}^a_{k-1,i}$ to $\mathbf{x}_{k,i}^b$. In particular, if there is no dynamical noise, then $\mathbf{x}_{k,i}^b = \mathcal{M}_{k,k-1} (\tilde{\mathbf{x}}^a_{k-1,i})$, with $\mathcal{M}_{k,k-1}$ being the transition operator (cf Section \ref{sec:ps}).
	
	END FOR
\item Filtering step: FOR i = 1 to N

	Multiply the weight $\tilde{w}_{k-1,i}$ of $\mathbf{x}_{k,i}^b$ by the likelihood $p(\mathbf{y}_k^o | \mathbf{x}^b_{k,i})$
	
	END FOR
	
	Apply Eq. (\ref{eq:weight_update_PF}) to obtain the normalized weights $\{ w_{k,i} \}_{i=1}^N$.
\item Re-sampling step:
	\begin{itemize}
	\item[--] Evaluate the difference $\delta_k$ between the weight ``entropy'' $- \sum\limits_{i=1}^{N} w_{k,i} \log(w_{k,i})$ and that with the uniform weight $1/N$, namely, $\delta_k = \log N + \sum\limits_{i=1}^{N} w_{k,i} \; \log(w_{k,i})$ \citep{Pham2001}
	\item[--] IF $\delta_k < 0.25$
	
	    \begin{myindentpar}{0.3cm}
    	No need to re-sample. Set $\tilde{\mathbf{x}}_{k,i}^a = \mathbf{x}_{k,i}^b$ and the associated weight $\tilde{w}_{k,i} = w_{k,i}$
    	\end{myindentpar}

	ELSE
		\begin{myindentpar}{0.3cm}
    	
    		FOR i = 1 to N
				\begin{itemize}
					\item[$\rhd$] Draw a sample $\tilde{\mathbf{x}}$ from the set $\{ \mathbf{x}_{k,i}^b, w_{k,i}\}_{i=1}^{N}$ through importance re-sampling, as in the 		
					bootstrap particle filter
					\item[$\rhd$] Draw a sample $\eta$ from the Gaussian pdf $N(\mathbf{0},\mathbf{I}_N)$
					\item[$\rhd$] Set $\tilde{\mathbf{x}}_{k,i}^a = \tilde{\mathbf{x}} + h \mathbf{S}_k \eta$ and the associated weight $\tilde{w}_{k,i} = 1/N$
					\item[$\rhd$] If desirable, introduce some additional ``jittering'' to $\tilde{\mathbf{x}}_{k,i}^a$
				\end{itemize}
			END FOR
    	\end{myindentpar}
    END IF 	
	\end{itemize}
	
\end{itemize}

\bibliographystyle{wileyqj}
\bibliography{./references}
\clearpage
\listoftables
\listoffigures
%
\clearpage
\begin{table*} 
\centering
\caption{\label{table:ess_of_normal_RPF} Time mean effective sample sizes (ESS) of the normal RPF and RPF-RN ($\beta = 0.02$) with different assimilation steps $S_a$ and observation noise variances $\gamma$ in the 1/2 observation scenario.}
\begin{tabular}{lllll}
\hline \hline
\multirow{2}{*}{RPF}& \multicolumn{4}{c}{$\gamma = $} \\
\cline{2-5}
& 0.01 & 0.1 & 1 & 10 \\
\hline
$S_a = 1$ & 1.0146 & 1.1631 & 3.1467 & 7.8407 \\
$S_a = 2$ & 10.4868 & 10.5510 & 11.2989 & 13.1758 \\
$S_a = 4$ & 15.2337 & 15.2572 & 15.5151 & 16.1476 \\
$S_a = 6$ & 16.8284 & 16.8407 & 16.9722 & 17.2499 \\
$S_a = 8$ & 17.6069 & 17.6145 & 17.6978 & 17.8570 \\
$S_a = 10$ & 18.0816 & 18.0863 & 18.1366 & 18.2112 \\
$S_a = 12$ & 18.4043 & 18.4085 & 18.4461 & 18.4994 \\
\hline \hline
\multirow{2}{*}{RPF-RN}& \multicolumn{4}{c}{$\gamma = $} \\
\cline{2-5}
& 0.01 & 0.1 & 1 & 10 \\
\hline
$S_a = 1$ & 1.0798  & 1.9776  & 5.9917  & 8.5888 \\
$S_a = 2$ & 10.5026 & 10.7161 & 12.3772 & 13.6230 \\
$S_a = 4$ & 15.2357 & 15.2831 & 15.7914 & 16.2794 \\
$S_a = 6$ & 16.8296 & 16.8490 & 17.0816 & 17.3037 \\
$S_a = 8$ & 17.6072 & 17.6189 & 17.7471 & 17.8508 \\
$S_a = 10$ & 18.0820 & 18.0903 & 18.1770 & 18.2380 \\
$S_a = 12$ & 18.4046 & 18.4111 & 18.4705 & 18.4992 \\
   \hline \hline
\end{tabular}
\end{table*}

\clearpage
\begin{table*} 
\centering
\caption{\label{table:time_mean_RMSE} Time mean RMSEs of the normal RPF and RPF-RN ($\beta = 1$) with (possibly) mis-specified forcing terms $F$ and the observation noise variances $\gamma$ in the 1/40 observation scenario.}
\begin{tabular}{lcccccc}
\hline \hline
\multirow{2}{*}{RPF}& \multicolumn{6}{c}{$\gamma = $} \\
\cline{2-7}
& 0.25 & 0.5 & 1 & 2 & 5 & 10 \\
\hline
$F = 4$ & 4.1448 & 4.0985  & 4.0616 & 3.9076 & 3.7926 & 3.7582 \\
$F = 6$ & 4.4318 & 4.2709  & 4.0929 & 3.8518 & 3.7481 & 3.6956 \\
$F = 8$ & 4.7182 & 4.3871  & 4.0555 & 3.8904 & 3.7821 & 3.7196 \\
$F = 10$ & 4.9694 & Div & Div & 3.9847 & 3.8579 & 3.8016 \\
$F = 12$ & Div & Div & Div & Div & Div & Div \\
\hline \hline
\multirow{2}{*}{RPF-RN} & \multicolumn{6}{c}{$\gamma = $} \\
\cline{2-7}
& 0.25 & 0.5 & 1 & 2 & 5 & 10 \\
\hline
$F = 4$ & 3.8406 & 3.8490  & 3.8556 & 3.8564 & 3.7957 & 3.7610 \\
$F = 6$ & 3.8707 & 3.9055  & 3.9096 & 3.8527 & 3.7576 & 3.7046 \\
$F = 8$ & 4.0495 & 4.0347  & 3.9758 & 3.8628 & 3.7636 & 3.7027 \\
$F = 10$ & 4.2397 & Div  & Div & 3.9477 & 3.8486 & 3.7975 \\
$F = 12$ & Div & Div & Div & Div & Div & Div \\
   \hline \hline
\end{tabular}
\end{table*}

%
\clearpage

\clearpage

\clearpage

\clearpage

\clearpage

\clearpage




\clearpage

\clearpage

\clearpage

\clearpage

\clearpage

\end{document}